\documentclass[prd,nofootinbib,twocolumn,superscriptaddress,preprintnumbers,longbibliography]{revtex4-1}

\usepackage{amsfonts, amssymb,amsmath,latexsym,graphics, graphicx,epsfig,multirow,comment,
,feyn,slashed,xcolor,afterpage, makecell} 
\usepackage{booktabs, blindtext}
\usepackage{tabularx,afterpage}
\usepackage[colorlinks=true
,urlcolor=blue
,anchorcolor=blue
,citecolor=blue
,filecolor=blue
,linkcolor=blue
,menucolor=blue
,linktocpage=true
,pdfproducer=medialab
,pdfa=true
]{hyperref}
\usepackage[utf8]{inputenc}
\usepackage{url}

\newcolumntype{C}[1]{>{\centering\let\newline\\\arraybackslash\hspace{0pt}}m{#1}}

\newcommand {\be} {\begin {equation}}
\newcommand {\ee} {\end {equation}} 

\newcommand {\bes} {\begin {equation*}}
\newcommand {\ees} {\end {equation*}}

\newcommand{\es}[2] {\begin{equation} \label{#1} \begin{split} #2 \end{split} \end{equation}}

\usepackage{csvsimple}

\newcommand{\beq}{\begin{equation}}
\newcommand{\eeq}{\end{equation}}

\begin{document}

\title{A Search for Dark Matter Annihilation in the Milky Way Halo}

\preprint{PUPT-2559}

\author{Laura J. Chang}
\affiliation{Department of Physics, Princeton University, Princeton, NJ 08544}

\author{Mariangela Lisanti}
\affiliation{Department of Physics, Princeton University, Princeton, NJ 08544}

\author{Siddharth Mishra-Sharma}
\affiliation{Department of Physics, Princeton University, Princeton, NJ 08544}

\date{\today}

\begin{abstract}
The Milky Way halo is the brightest source of dark matter annihilation on the sky.  Indeed, the potential strength of the Galactic dark matter signal can supersede that expected from dwarf galaxies and galaxy groups even in regions away from the Inner Galaxy.  In this paper, we present the results of a search for dark matter annihilation in the smooth Milky Way halo for $|b| > 20^\circ$ and $r < 50^\circ$ using 413 weeks of \emph{Fermi} Pass~8~data within the energy range of $\sim$0.8--50~GeV.  We exclude thermal dark matter with mass below $\sim$70~GeV that annihilates to $b\bar{b}$ at the 95\% confidence level using the  \texttt{p6v11} cosmic-ray foreground model, providing the strongest limits on the annihilation cross section in this mass range.  These results exclude the region of dark matter parameter space that is consistent with the excess of $\sim$GeV photons observed at the Galactic Center for the $b\bar{b}$ annihilation channel and, for the first time, start probing the $\tau^+\tau^-$ explanation. We explore how these results depend on uncertainties in the foregrounds by varying over a set of reasonable models.
\end{abstract}
\maketitle

\section{Introduction}  
\label{sec:intro}

The \emph{Fermi} Large Area Telescope~\cite{Atwood:2009ez} provides an unprecedented view of the gamma-ray sky.  The all-sky maps that are available can harbor clues about the nature of dark matter (DM), which can annihilate to visible states that produce showers of high-energy photons.  A variety of such searches have been performed, focusing on regions where the relative DM density is expected to be significant.  Thus far, the most sensitive bounds come from looking at ultrafaint dwarf galaxies~\cite{Ackermann:2013yva,Geringer-Sameth:2014qqa,Ackermann:2015zua,Fermi-LAT:2016uux} and galaxy groups~\cite{Lisanti:2017qoz,Lisanti:2017qlb}.  In this paper, we explore emission due to annihilating DM from the Galactic halo, and demonstrate that it can be used to set robust constraints on the DM annihilation cross section.  These constraints are the strongest to date on DM with mass less than $\sim$70~GeV, for the $b\bar{b}$ annihilation benchmark. 

The halo surrounding our Galaxy provides the brightest source of DM emission on the sky.  In general, the DM flux is proportional to the so-called $J$-factor, which is the integral over the line-of-sight, $s$, and solid angle, $\Omega$, of the squared DM density profile:
\begin{equation}
J = \int ds \, d\Omega \, \rho^2(s,\Omega) \,. 
\label{eq:Jfactor}
\end{equation}
The $J$-factor provides a useful metric for comparing the strength of an annihilation signal expected from different targets.  For example, the $J$-factors from some of the brightest ultrafaint dwarf galaxies are $\sim$$10^{19}$~GeV$^2$\,cm$^{-5}$\,sr~\cite{Fermi-LAT:2016uux}, comparable to those of the brightest galaxy groups~\cite{Lisanti:2017qoz}.
In contrast, the center of our own Galaxy has a $J$-factor several orders of magnitude larger, with $J$$\sim$$10^{23}$~GeV$^2$\,cm$^{-5}$\,sr 
 in the inner $40^\circ\times40^\circ$ region.  Even if one were to avoid the central part of the Galaxy and only consider an annulus of $r < 50^\circ$ and latitudes greater than $|b| > 20^\circ$, the $J$-factor is still as large as $\sim$$10^{22}$~GeV$^2$\,cm$^{-5}$\,sr.

Despite the strength of the smooth Galactic DM signal, many other factors complicate a potential search.  The primary challenge is posed by the bright diffuse emission from cosmic rays propagating in the Galaxy.  These contributions arise from  $\pi^0$ decay, Bremsstrahlung from the interaction of cosmic-ray electrons with interstellar gas, and inverse-Compton (IC)  scattering of photons off of high-energy electrons.  This diffuse foreground contributes the vast majority of the high-energy photons we see on the sky, accounting for $\sim$50--90\% of the observed photons depending on the energy range considered~\cite{Ackermann:2014usa}, and is challenging to model accurately. Any search for Galactic DM must mitigate these uncertainties and quantify the effects of varying over assumptions in the foreground models.

Searches for Galactic DM can be divided into two broad categories.  The first set focuses on the Inner Galaxy, within $r\lesssim 20^\circ$~\cite{Goodenough:2009gk, Hooper:2010mq, Boyarsky:2010dr, Hooper:2011ti, Abazajian:2012pn, Hooper:2013rwa, Gordon:2013vta, Huang:2013pda, Macias:2013vya, Abazajian:2014fta, Daylan:2014rsa, Zhou:2014lva, Calore:2014xka, Abazajian:2014hsa, TheFermi-LAT:2015kwa, Karwin:2016tsw}.  These analyses have conclusively found an excess of $\sim$GeV photons whose energy distribution and spatial morphology can be consistent with the expectation due to DM annihilation.  However, recent studies have shown that the distribution of photons in the Inner Galaxy is more consistent with a population of unresolved point sources, disfavoring the DM interpretation~\cite{Lee:2015fea, Bartels:2015aea}.  Additionally, other studies  suggest that the spatial morphology of the excess may better trace the stellar bulge~\cite{Macias:2016nev, Bartels:2017vsx}.  Complementary studies of Milky Way dwarfs~\cite{Fermi-LAT:2016uux} and galaxy groups~\cite{Lisanti:2017qoz,Lisanti:2017qlb} are starting to put in tension the DM interpretation of the excess emission. However, the tension can be alleviated depending on the specific assumptions made about, \emph{e.g.}, the dwarf halo profiles~\cite{Keeley:2017fbz,Geringer-Sameth:2014qqa,Sanders:2016eie}; the stellar membership criteria used to infer the dwarf halo properties~\cite{2016MNRAS.462..223B,Geringer-Sameth:2014yza}; the shape of the Milky Way halo~\cite{Abazajian:2015raa}; or the nature of substructure boost in galaxy groups~\cite{Lisanti:2017qlb}.

Even though the Galactic Center is the brightest DM source on the sky, it is also one of the most complicated due to the large astrophysical foregrounds. A complementary approach to looking at the Inner Galaxy relies on looking at the Galactic halo at higher latitudes where the DM density is still large, but the foreground levels are much smaller~\cite{Papucci:2009gd, Cirelli:2009dv, Baxter:2011rc, Malyshev:2010zzc,Ackermann:2012rg, Zechlin:2017uzo,Huang:2015rlu}. This is the approach that we take in this work.  Focusing on a region defined by $|b| > 20^\circ$ and $r < 50^\circ$, we search for signals of DM annihilation from the smooth Milky Way halo.  
The limits obtained provide the strongest constraints on low-mass DM annihilation signals and tightly constrain the DM interpretation of the GeV Excess. We verify the robustness of these results in the presence of a potential DM signal and discuss how they are affected by variations in the Galactic foreground models. \vspace{0.02in}

This paper is organized as follows. In Sec.~\ref{sec:analysis} we describe our analysis pipeline and the statistical procedure employed. In Sec.~\ref{sec:results} we present the results of our study, discussing the effects of Galactic foreground mismodeling and steps taken to reduce their impact.  We also discuss the implications of our results for the DM interpretation of the Galactic Center Excess. We conclude in Sec.~\ref{sec:conclusions}.

\section{Analysis Procedure}
\label{sec:analysis}

We make use of 413 weeks of {\it Fermi}-LAT Pass 8 data collected between August 4, 2008 and July 7, 2016. We analyze the subset of photons in the \texttt{ULTRACLEANVETO} event class, restricting to the top quarter of photons by quality of point-spread function (PSF) reconstruction (corresponding to PSF3 event type). The data is binned in 18 logarithmically-spaced energy bins between $\sim$0.8--50~GeV.  The recommended quality cuts are applied, corresponding to zenith angle less than $90^\circ$, $\texttt{LAT\_CONFIG}=1$, and $\texttt{DATA\_QUAL}>0$.\footnote{\url{https://fermi.gsfc.nasa.gov/ssc/data/analysis/documentation/Cicerone/Cicerone_Data_Exploration/Data_preparation.html}}
Each energy bin is spatially binned into individual pixels using~\texttt{HEALPix}~\cite{Gorski:2004by} with \texttt{nside}\,=\,128; the dataset is thus reduced to an array of integers that describes the number of photons in the energy bin, $i$, and pixel, $p$.  

Template fitting is a standard astrophysical procedure where the data is described by a set of spatial maps (referred to as templates) that are binned in the same way as the data, which describe the separate components that contribute to the total photon count.  Each template is associated with a normalization that is treated as a free model parameter in the fit.  The likelihood for a given energy bin is then a product of the Poisson probabilities associated with the observed counts $n_i^{p}$ in each pixel of the region-of-interest:  
\begin{equation}
\mathcal{L}_i(d_i | {\boldsymbol \theta}_i) = \prod_p \frac{\mu_i^{p}({\boldsymbol \theta}_i)^{n_i^{p}} e^{-\mu_i^{p}({\boldsymbol \theta}_i)}}{n_i^{p}!}\,,
\label{eq:pi}
\end{equation}
where $d_i$ denotes the data in energy bin $i$, ${\boldsymbol \theta_i}$ represents the set of model parameters and $\mu_i^{p}({\boldsymbol \theta}_i)$ is the number of expected counts in a given pixel and energy bin.  The total likelihood is simply the product over the individual $\mathcal{L}_i$ for each energy bin.    

The region-of-interest (ROI) for this study is chosen to maximize the strength of the DM signal while minimizing the effects of foreground mis-modeling.  Specifically, we take $|b| > 20^\circ$ to avoid the Galactic plane, where cosmic-ray emission is particularly bright and there are more unresolved point sources.  In addition, we take $r < 50^\circ$ to reduce the possibility of over-subtraction and/or spurious excesses obtained from fitting the foreground model over large sky areas.  Better modeling of the Galactic diffuse emission is an ongoing effort~\cite{Gaggero:2015nsa,Werner:2014sya,Kissmann:2014sia,Gaggero:2014xla,Evoli:2016xgn}, and the use of image reconstruction and parametric modeling techniques such as \textsc{SkyFACT}~\cite{Storm:2017arh} and D$^3$PO~\cite{Selig:2014qqa,Huang:2015rlu} could improve modeling of the Galactic diffuse emission over significantly larger regions of the sky.  Increasing the radial cut in $r$ beyond that used here could potentially improve sensitivities to DM by $\sim$20--30\% or more, depending on the density profile---see App.~\ref{sec:roi} for a discussion of Asimov projections.

The expected photon count, $\mu_i^{p}({\boldsymbol \theta}_i)$, in each pixel of the ROI depends on contributions from standard astrophysical sources as well as DM, if present.  We account for four astrophysical components that trace: (i)~the Galactic diffuse emission, as described by the  {\it Fermi} \texttt{gll\_iem\_v02\_P6\_V11\_DIFFUSE (p6v11)} model,\footnote{\url{https://fermi.gsfc.nasa.gov/ssc/data/access/lat/ring_for_FSSC_final4.pdf}} (ii)~the {\it Fermi} bubbles~\cite{Su:2010qj}, (iii)~isotropic emission, and (iv)~\emph{Fermi} 3FGL point sources~\cite{Acero:2015hja}.  The smooth Galactic DM template is modeled using a generalized Navarro-Frenk-White (NFW) profile~\cite{Navarro:1996gj}:
\begin{equation}
\rho_\mathrm{NFW}(r) = \frac{\rho_0}{(r/r_s)^\gamma[1+(r/r_s)]^{3-\gamma}}
\label{eq:NFW}
\end{equation} 
with inner-slope $\gamma=1$, scale radius $r_s = 17$ kpc, and local density $\rho(r_\odot) = 0.4$ GeV$\,$cm$^{-3}$~\cite{2015ApJ...814...13M,Sivertsson:2017rkp} at the Solar position $r_\odot = 8$~kpc~\cite{Read:2014qva}.  All templates are smoothed with the energy-dependent PSF of the LAT instrument, modeled as a King function.\footnote{\url{https://fermi.gsfc.nasa.gov/ssc/data/analysis/documentation/Cicerone/Cicerone_LAT_IRFs/IRF_PSF.html}}

In our fiducial study, we use the \texttt{p6v11} Galactic diffuse emission model, which is designed to capture changes in the cosmic-ray emission on the full sky as a function of Galactocentric radius.  The model includes contributions from $\pi^0$-decay and Bremsstrahlung emission, as traced by maps of gas column-densities, as well as inverse-Compton emission, as predicted using \textsc{Galprop}~\cite{Strong:1998fr}; the relative normalizations of these separate components are fixed.  The fact that the \texttt{p6v11} model should be used with caution for energies above $\sim$50~GeV sets our upper energy cut-off.  To give the \texttt{p6v11} template more freedom, we divide it into eight radial slices of equal area within our ROI.  The normalization of each slice is then varied separately in the fitting procedure. Each slice is roughly $\sim$440~deg$^2$ in area, comparable in size to the regions used in dwarf and galaxy group studies ($\sim$100 and 316 deg$^2$, respectively)~\cite{Fermi-LAT:2016uux,Lisanti:2017qlb} and smaller than the typical regions used in Inner Galaxy analyses ($\sim$1600 deg$^2$)~\cite{Daylan:2014rsa,Calore:2014xka}. The additional freedom given to the Galactic diffuse template allows the fit to better account for localized excesses or mis-modeled features in the emission. Note that we do not use the \texttt{gll\_iem\_v06 (p8R2)} model~\cite{Acero:2016qlg}, which is recommended for the Pass~8 dataset.  \texttt{p8R2} includes large-scale residuals obtained from a fit to the \emph{Fermi} data that have been added back into the model, and is therefore not appropriate when searching for extended DM signals. We also do not use the \texttt{p7v6} diffuse model, which contains large-scale structures including Loop~I and the \emph{Fermi} bubbles with a fixed normalization.

The \texttt{p6v11} model does not include known large-scale structures that overlap with the ROI, such as the \emph{Fermi} bubbles.  We account for the bubbles by adding two templates that model the Northern and Southern lobes.  The shape of the lobes is inferred directly from \emph{Fermi} data~\cite{Su:2010qj}, and the intensity of the emission is taken to be flat.  We let the normalization of the Northern and Southern lobes float independently in the fit. The ROI also overlaps with Loop~I, a large radio lobe in the Northern hemisphere~\cite{1962MNRAS.124..405L, 1981A&A...100..209H}.   While features corresponding to the radio observations have been observed in the~\emph{Fermi} data~\cite{2009arXiv0912.3478C, Su:2010qj, Ackermann:2012uf, Fermi-LAT:2014sfa}, significant uncertainties remain in the modeling of the spatial and intensity profile of Loop~I in gamma rays.  As a result, we conservatively do not include a template that traces Loop~I in our fiducial study.  We have performed variants of the fiducial study to assess the impact of this choice.  We find that the inclusion of an additional isotropic template in the Northern hemisphere as a proxy for Loop~I emission strengthens the limit by a factor of $\lesssim1.2$.

The isotropic template is intended to primarily capture extragalactic gamma-ray emission from unresolved sources such as blazars and star-forming galaxies, as well as more exotic contributions from extragalactic DM annihilation.  The inclusion of the point-source template accounts for emission from resolved (Galactic and extragalactic) sources.  The normalizations of all the sources are floated together in the template after fixing their individual fluxes to the values predicted by the 3FGL catalog.  We note that all 3FGL sources are conservatively masked to 95\% containment in PSF for the corresponding energy bins.  Therefore, the primary purpose of the point-source template is to account for any potential mis-modeling in the tails of the emission.
  
To summarize, there are twelve free parameters associated with the astrophysical components---eight for the Galactic diffuse slices, two for the \emph{Fermi} bubbles, and one each for the isotropic and point-source templates.  As we are ultimately interested in the intensity of the DM signal, we treat these as nuisance parameters and remove them using the profile likelihood method~\cite{Rolke:2004mj}.  Specifically, we build a likelihood profile for the intensity associated with DM annihilation in the smooth Galactic halo, fixing the normalization of this template at various values while profiling over the astrophysical components.  The resulting likelihood  only depends on the DM intensity in each energy bin, which is related to the annihilation cross section, $\langle \sigma v \rangle$, and mass, $m_{\chi}$, through the expression for the differential gamma-ray flux:
\begin{equation}
\frac{d\Phi}{dE_{\gamma}} =   J \, \times \frac{\langle\sigma v\rangle}{8 \pi m_{\chi}^{2}} \, \, \sum_j \text{Br}_{j}\, \frac{dN_{j}}{dE_{\gamma}}   \,,
\end{equation}
where $E_\gamma$ is the gamma-ray energy and $\text{Br}_{j}$ is the branching fraction to the $j^\text{th}$ annihilation channel.  The energy spectrum for each channel is described by the function $dN_{j}/dE_{\gamma}$, which is modeled using \texttt{PPPC4DMID}~\cite{Cirelli:2010xx}.  Note that we do not account for DM substructure in the Milky Way halo in this study, which would increase the strength of the annihilation signal.  Given the theoretical uncertainties associated with modeling the spatial distribution and properties of DM subhalos, such a search deserves its own dedicated study. 

The test statistic (TS) profile for $\langle \sigma v \rangle$ is  defined as 
\es{TS}{
{\rm TS} \equiv 2  \left[ \log \mathcal{L}(d | \mathcal{M}, \langle\sigma v\rangle, m_\chi ) -  \log \mathcal{L}(d | \mathcal{M}, \widehat{\langle\sigma v\rangle}, m_\chi ) \right] \,, 
}
where $\widehat{\langle \sigma v\rangle}$ is the cross section that maximizes the likelihood for a specified DM model, $\mathcal{M}$, of given annihilation channel and mass.  The TS is nonpositive by definition and can be
used to set a threshold for limits on the cross section. In particular, the 95\% upper limit on the annihilation cross section is given by the value of $\langle \sigma v \rangle$ associated with $\text{TS}=-2.71$.  We implement template fitting with the package \texttt{NPTFit} \cite{Mishra-Sharma:2016gis} and use the L-BFGS-B~\cite{citeulike:10176711} minimization algorithm implemented through \texttt{SciPy}~\cite{scipycite}.

\begin{figure}[t]
\centering
\includegraphics[width=0.49\textwidth]{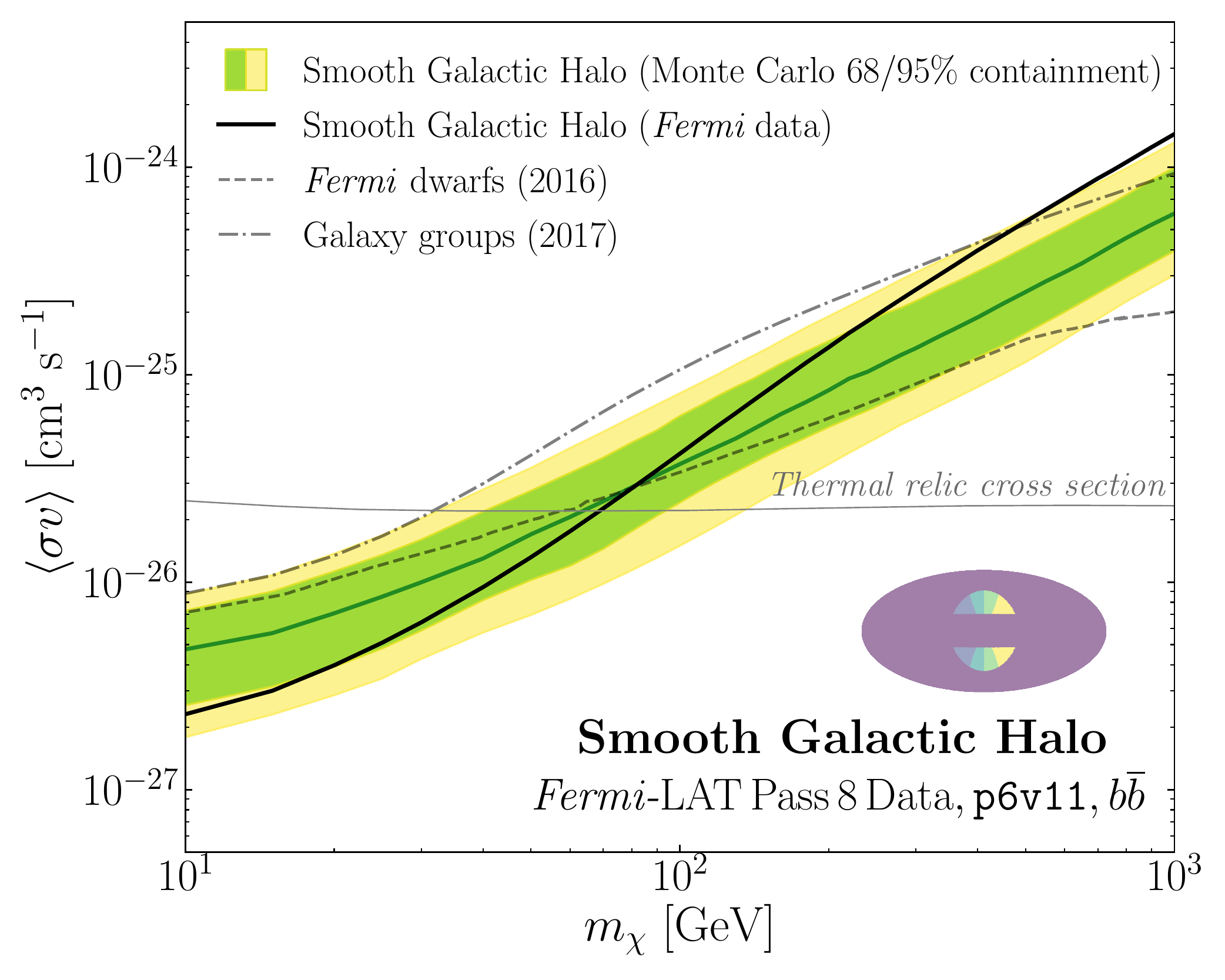} 

\caption{The solid black line shows the 95\% confidence limit on dark matter of mass, $m_\chi$, annihilating with cross section, $\langle \sigma v \rangle$, in the smooth Galactic halo, within $|b| > 20^\circ$ and $r < 50^\circ$, obtained using the \texttt{p6v11} foreground model.  The green(yellow) band denotes the 68(95)\% containment for the expected sensitivity, as derived from Monte Carlo simulations.  For the Galactic halo, we assume a generalized NFW profile with inner slope of $\gamma = 1$ and local density $\rho(r_\odot) = 0.4$~GeV~cm$^{-3}$.  We also show the corresponding limits obtained from dwarf galaxies~\cite{Fermi-LAT:2016uux} and galaxy groups~\cite{Lisanti:2017qlb} (grey dashed and dot-dashed lines, respectively).  The expected annihilation cross section for a generic weakly interacting massive particle is indicated by the solid grey line~\cite{Steigman:2012nb}. The inset depicts the eight radially sliced regions within the fiducial ROI over which the \texttt{p6v11} template is allowed to float.}
\label{fig:bounds}
\end{figure}

We have performed numerous tests to ensure that the statistical procedure outlined above can recover a potential signal in the data.  Such tests are crucial in verifying the robustness of these methods, especially given the potentially large degeneracies between the signal and foreground components, which are both diffuse in nature.  Additionally, the freedom given to the foreground emission by separately fitting its normalization in the radial slices can lead to challenges in regimes of low photon statistics.  We have performed tests on both data and Monte Carlo and verified that our analysis procedure would not exclude a DM signal if one were present in the data. A detailed description of these tests is provided in App.~\ref{sec:siginj}.  \vspace{0.02in}

\begin{figure*}[t]
\centering
\includegraphics[width=0.91\textwidth]{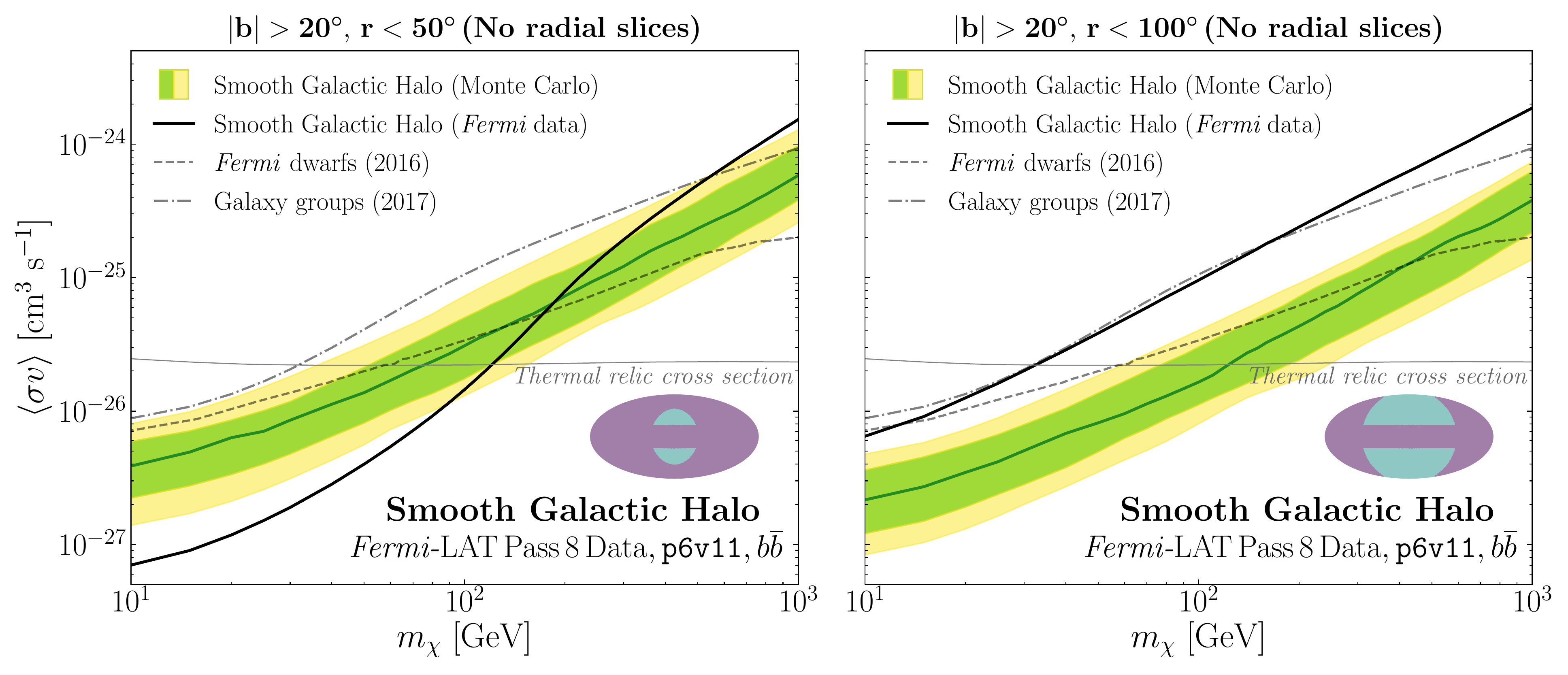} \hspace{4mm}
\caption{The limit when the  \texttt{p6v11} template is not divided into eight radial slices whose normalizations float independently in the fitting procedure. For each panel, the inset depicts the regions (denoted in turquoise) over which the \texttt{p6v11} template is allowed to float. A region corresponding to $|b| > 20^\circ$ and $r < 50^\circ$($100^\circ$) is used in the left(right) panel.}
\label{fig:roiselectionp6}
\end{figure*}

\section{Results and Discussion}
\label{sec:results}

\subsection{Dark Matter Annihilation Limit}
\label{subsec:limit}

Figure~\ref{fig:bounds} shows the 95\% confidence limit on the DM annihilation cross section into the $b\bar{b}$ final state (solid black).  For comparison, the published limits from the most recent dwarf~\cite{Fermi-LAT:2016uux} and galaxy group~\cite{Lisanti:2017qlb} studies appear as the grey dashed and dot-dashed lines,  respectively.  The $b\bar{b}$ limits from the smooth Galactic halo are the strongest to date for DM masses below $\sim$70~GeV.

The green(yellow) band in Fig.~\ref{fig:bounds} shows the 68(95)\% expected sensitivity obtained from Monte Carlo simulations.  To make the simulations, we Poisson fluctuate the sum of best-fit templates on data within the ROI, letting the normalizations for the different foreground slices and bubble lobes float independently.  The sensitivity projection is derived from 100 Monte Carlo variations. A data-driven foreground expectation obtained  by looking at a large number of blank fields, as is standard for dwarf and galaxy group studies, is not feasible for Galactic DM searches because the overall size of the ROI is a substantial fraction of the full sky.  The Monte Carlo bands do, however, provide an important comparison benchmark.  For example, if the Galactic foregrounds are over-subtracted in the fitting procedure, then the data limits will be artificially strengthened and appear stronger than the Monte Carlo expectation. 

While the morphology of the signal template suggests that one should minimize the latitude cut ($|b| > b_\text{cut}$) and maximize the radial cut ($r < r_\text{cut}$) for optimal sensitivity to DM (see App.~\ref{sec:roi} for more details), a full-sky analysis is not viable in actuality due to the large uncertainties associated with modeling the Galactic foregrounds.  As a result, we conservatively choose $b_\text{cut}= 20^\circ$ to avoid the Galactic plane, where the foregrounds are particularly bright and there is increased contamination from unresolved point sources.  In addition, we choose $r_\text{cut} = 50^\circ$ because fitting over larger sky regions can lead to over-subtraction and/or spurious excesses in the data analysis. While the definition of the fiducial ROI is intended to mitigate the large systematic uncertainties associated with the foregrounds, we also give the \texttt{p6v11} template additional freedom by fitting its normalization separately in eight radial slices of equal area, as discussed in Sec.~\ref{sec:analysis}. Figure~\ref{fig:roiselectionp6} demonstrates the need for these additional steps. The left panel shows the data limit and corresponding Monte Carlo expectation obtained when the \texttt{p6v11} template is not divided into eighths, for our fiducial ROI. The right panel shows the case corresponding to a larger radial cut $r_\text{cut} = 100^\circ$. Every other aspect of the analysis is kept the same as in the fiducial study in these cases, except that the Northern and Southern lobes of the \emph{Fermi} bubbles are floated together.\footnote{Doing the same for the fiducial study does not change the result.} The projected sensitivities obtained from Monte Carlo simulations are essentially equivalent between the fiducial study and these two examples. The data limits, on the other hand, are starkly different. A large excess in the data limit compared to the Monte Carlo expectation is apparent when the larger ROI is used. When the fiducal ROI is used but the foreground template is not broken into radial slices, over-subtraction leads to artificially strong bounds.  We therefore conclude that performing the fit over smaller sky regions and varying the \texttt{p6v11} template over additional degrees of freedom stabilizes the analysis in the designated ROI. 
\vspace{0.02in}

\subsection{Galactic Foreground Modeling}
\label{sec:foreground}

Uncertainties due to modeling of the Galactic diffuse emission are inherent in searches for large-scale gamma-ray structures. We have made an effort to minimize the effects of these uncertainties by giving more degrees of freedom to the \texttt{p6v11} template.  However, inherent assumptions that go into the construction of the template can still have a potentially large effect on the final result.  Here, we present results for three additional foreground models that are designed to span several well-motivated possibilities.   Our approach is to understand  how each set of assumptions regarding the cosmic-ray modeling impacts the DM sensitivity for the ROI considered in this work.  

We repeat the analysis using Models A, B, and C, which were developed by the \emph{Fermi}-LAT Collaboration specifically for their study of the isotropic gamma-ray background at higher latitudes~\cite{Ackermann:2014usa}. These models make distinct but well-motivated choices for the cosmic-ray source distribution, diffusion coefficients, and re-acceleration strengths that span a wide range of possibilities.  Separate templates for $\pi^0$ decay, Bremsstrahlung, and inverse-Compton (IC) emission are provided, so their normalizations can be varied independently in the fitting procedure.  In these analyses, we use a single combination of the Bremsstrahlung and $\pi^0$-decay templates as obtained from a fit to data using eight separate equal-area slices. Both these components trace the diffuse gas and dust structures in the Galaxy, so giving them separate degrees of freedom is expected to have a negligible effect on the results.

We highlight the fact that the IC and $\pi^0$+Bremsstrahlung templates are allowed to vary separately in the Model A, B, and C fits.  As a result, the foreground templates in these tests are given considerable freedom in the fitting procedure, as they are associated with sixteen free parameters (rather than just eight, as in the \texttt{p6v11} case).  This is a very important cross-check of the fiducial results, because the relative normalizations of foreground components are fixed in \texttt{p6v11}, with the ratio set by a previous fit to the data.  However, because that fit did not include a DM template, one might worry that a potential signal---if present---would be absorbed by the foreground components (particularly the IC component) in the initial fitting procedure. If this were the case, using \texttt{p6v11} for a Galactic DM search could potentially give artificially stringent DM limits.  

\begin{figure*}[t]
\centering
\includegraphics[width=\textwidth]{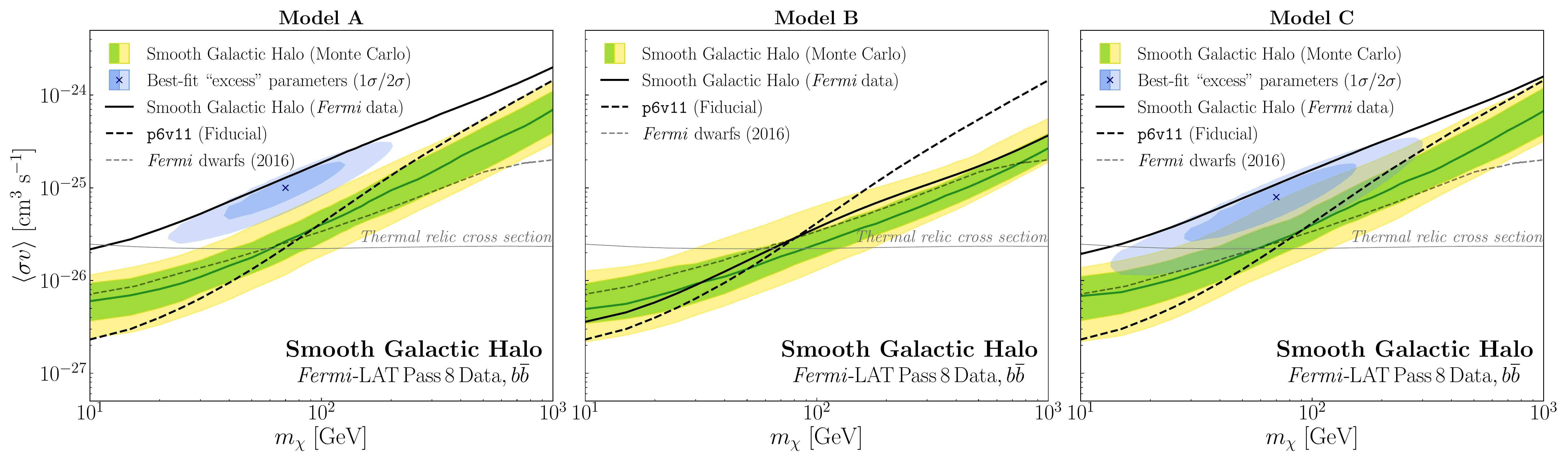}

\caption{Similar to Fig.~\ref{fig:bounds} of the paper, except using the Model~A, B, and C foreground models (left, middle, and right panel, respectively) as provided by the \emph{Fermi}-LAT Collaboration~\cite{Ackermann:2014usa}.  Note that the foreground templates are still divided into eight radial slices, as in the fiducial study, but the normalizations of the inverse-Compton and $\pi^0$+Bremsstrahlung templates are allowed to float independently.  The fiducial limit obtained using the \texttt{p6v11} foreground model is shown by the dashed black line for comparison.  The ``excesses'' in the Model~A and C studies (with significances $\mathrm{TS}_\mathrm{max}\sim28$ and 14, respectively) are well-understood in terms of the source populations included in these models; see text for further discussion. Model~B is statistically preferred over Models A and C as a description of the data in our ROI;  the difference in the maximum log-likelihood between Model~B and~A(C) is $\Delta \log\mathcal{L}_\text{max} = 136(119)$.}
\label{fig:modelABC}
\end{figure*}

Fig.~\ref{fig:modelABC} shows the limits obtained using Models~A, B, and C.  The differences between the results can be understood in terms of the assumptions going into the separate models, which we now describe in detail: 

Model~A is based on the class of Galactic diffuse models studied in~\cite{2012ApJ...750....3A}, and is described in detail in~\cite{Ackermann:2014usa}.  Here, we only highlight the main elements that distinguish it from Models~B and~C.  For Model~A, cosmic-ray electrons and nuclei are both sourced by the same population of pulsars, and the cosmic-ray diffusion coefficient and re-acceleration strength are held constant.   The left panel of Fig.~\ref{fig:modelABC} shows the Monte Carlo expectation and data limit when rerunning the fiducial analysis using the Model~A templates.  The recovered data limit is  weaker than the Monte Carlo expectation, which suggests that there is excess gamma-ray emission in the ROI that is not captured by the Model~A templates.  It should be noted that the foreground templates are given considerable freedom in the fitting procedure, as the normalizations of the $\pi^0$+Bremsstrahlung and IC templates are allowed to float separately in each radial slice.  Despite this freedom, a large amount of DM emission is still needed to improve the quality of the fit. A DM ``excess'' with a $\mathrm{TS}_\mathrm{max}\sim28$ is observed, with the best-fit 1$\sigma$ and 2$\sigma$ (corresponding to deviations in TS of $-2.30$ and $-6.18$ from the global maximum) containment regions as shown in the figure. The fact that the DM parameter space that is favored is clearly excluded by the dwarf searches strongly suggests that the weakening of the bounds is not due to DM, and is likely of astrophysical origin.

Model~B provides an important counterpoint to Model~A ~\cite{Ackermann:2014usa}.  It includes an additional source population of electrons at the Galactic Center, which contributes to the IC emission.  Unlike Model~A, which closely reproduces the local cosmic-ray electron spectrum, Model~B under-predicts the distribution below $\sim$20~GV. However, this disparity can be accounted for by contributions from other more local sources. The middle panel of Fig.~\ref{fig:modelABC} shows the Monte Carlo expectation and data limit for the Model~B study.  The limit is comparable to the fiducial case at low masses and is somewhat tighter for masses above $\sim$100~GeV, although still consistent within the Monte Carlo expectation. The predicted IC spectrum from \texttt{Galprop} that is used in Model~B tends to be a better match to the fitted spectrum (compared to Models A and C).  The better overall fit of Model B to the data in this case and the fact that the additional emission is absorbed by the IC template means that an astrophysical origin of the excess is statistically preferred to the DM component.

For Model~C, the cosmic-ray diffusion coefficient and re-acceleration strength depends on the Galactocentric radius and height~\cite{Ackermann:2014usa}.  Additionally, while the cosmic-ray electron/nuclei are sourced from the same population, their distribution is more central than that used for Model~A.  The differences between Model A and C predominantly show up in the outer galaxy, and so the two give largely similar results when used within our ROI. The excess emission observed in the case of Model~A is also present using Model~C, with a preference for roughly similar DM parameter values. 
Again, the fact that the preferred parameter space is robustly ruled out by dwarf searches strongly indicates that the excess emission in this case is of astrophysical origin.

To summarize, Model B provides limits very similar to those obtained in the fiducial case, while Models A and C exhibit significant excesses above Monte Carlo expectation. This difference can be attributed to the fact that Model~B includes an additional population of electron-only sources near the Galactic Center that contributes to the IC emission. 
Omission of this population in Models A and~C causes the DM template to absorb more flux, thus weakening the overall bounds. Overall, the fitted IC normalization for Model~B is closer to its initial \texttt{Galprop} prediction (with a value $\sim$1.1), as compared to that for Models~A and~C (with a value $\sim$2.4)~\cite{Ackermann:2014usa}. This suggests that, of the three scenarios considered, Model~B may best capture the IC emission in the ROI used here.
\begin{figure*}[htbp]
\centering
\includegraphics[width=\textwidth]{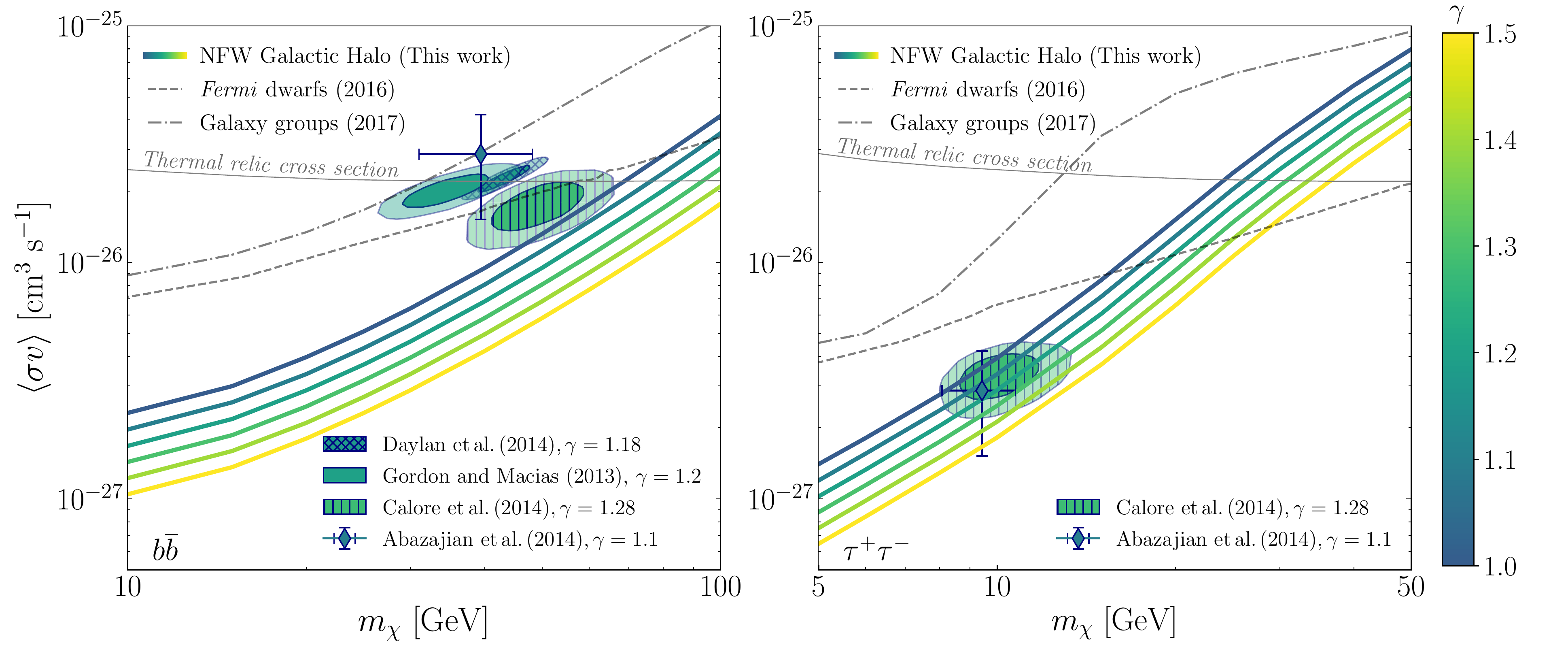} 

\caption{The 95\% confidence limits on dark matter annihilation into $b\bar{b}$ (left) and $\tau^+\tau^-$ (right) for the fiducial analysis, varying over the inner slope, $\gamma$, of the generalized NFW density profile.  The limits tighten as $\gamma$ increases; the lines shown correspond to linearly spaced steps from  $\gamma = 1$ to $1.5$.  The best-fit parameters obtained by previous studies of the GeV excess are indicated by the data point~\cite{Abazajian:2014fta} and the solid~\cite{Gordon:2013vta}, cross-hatched~\cite{Daylan:2014rsa}, and hatched~\cite{Calore:2014xka} regions.  Each region is indicated by 1$\sigma$/2$\sigma$ contours and colored corresponding to the best-fit $\gamma$ obtained by that study, also specified in the legend.  For ease of comparison, we have rescaled the best-fit cross-sections to be consistent with $\rho(r_\odot) = 0.4$ GeV\,cm$^{-3}$.   The corresponding limits obtained from dwarf galaxies~\cite{Fermi-LAT:2016uux} and galaxy groups~\cite{Lisanti:2017qlb} (grey dashed and dot-dashed lines, respectively) are also shown.  The expected annihilation cross section for a generic weakly interacting massive particle is indicated by the solid grey line~\cite{Steigman:2012nb}.}
\label{fig:gce_bounds}
\end{figure*}

\subsection{The GeV Excess}
\label{subsec:gce}

The results presented in this paper have direct implications for the interpretation of the excess of GeV photons observed in the Galactic Center.  If the GeV excess arises from DM, then the signal should also contribute a photon flux in the ROI studied here.  This is a more direct comparison than using dwarf galaxies or galaxy groups because it removes uncertainties having to do with differences in halo density distribution. In Fig.~\ref{fig:gce_bounds}, we show the Galactic DM limits obtained for various assumptions of the inner slope, $\gamma$, of the generalized NFW density profile.  The steeper the inner slope, the stronger the annihilation limit. Results are shown for annihilation into $b\bar b$ (left) and $\tau^+\tau^-$ (right). For comparison, we also show the best-fit regions to the Galactic Center gamma-ray excess from previous work as the data point~\cite{Abazajian:2014fta} and solid~\cite{Gordon:2013vta}, cross-hatched~\cite{Daylan:2014rsa}, and hatched~\cite{Calore:2014xka} regions. The DM interpretation of the GeV excess typically prefers a steeper inner slope with $\gamma \gtrsim 1.1$, where the limits from the Galactic halo become quite stringent.  These Milky Way limits robustly exclude a DM interpretation of the excess for the $b\bar b$ channel, and, for the first time, start probing the $\tau^+\tau^-$ scenario.  Explanations in terms of other annihilation channels are also highly constrained, as reviewed in App.~\ref{sec:extended}.\vspace{0.02in}

Variations in the foreground modeling can affect the recovered limits from our analysis and their implications for the  Galactic Center Excess.  Of the variations explored in Sec~\ref{sec:foreground}, Model~B appears to best capture the IC emission in the ROI, as the fitted normalization of this component is closest to its initial \texttt{Galprop} value.  The limits obtained using Model~B are only marginally weaker than those using \texttt{p6v11} at low masses and still robustly disfavor the DM interpretation of the excess in terms of annihilation into the $b\bar b$ final state.  These results are suggestive, but do not eliminate the systematic uncertainties associated with diffuse emission modeling.  To sidestep this issue, we can choose to compare our results to only those Inner Galaxy studies that use the same Galactic foreground model as we do.  The cross-hatched region in Fig.~\ref{fig:gce_bounds} is derived using the \texttt{p6v11} diffuse foreground model~\cite{Daylan:2014rsa} and therefore provides the most direct comparison to our limit.  It is strongly excluded by the limit we recover for the corresponding value of $\gamma$.  

\section{Conclusions}
\label{sec:conclusions}

In this paper, we have presented a comprehensive search for DM annihilation from the smooth Milky Way halo in \emph{Fermi} gamma-ray data. We do not find significant evidence for an annihilation signal, and obtain strong bounds on the properties of annihilating DM. We exclude thermal dark matter at masses below $\sim$70~GeV for the $b\bar{b}$ annihilation channel when using the \emph{Fermi} \texttt{p6v11} diffuse model, representing the strongest limits to date in this mass range. We have carefully considered uncertainties associated with the modeling of the diffuse Galactic foregrounds and are able to understand these variations in terms of the different physical assumptions underlying the foreground models. We have performed rigorous Monte Carlo and injected signal tests to ensure the robustness of our results.  This study excludes the $b\bar{b}$ annihilation interpretation of the Galactic Center excess at 95\% confidence for the \texttt{p6v11} diffuse model, and for the first time starts probing the $\tau^+\tau^-$ annihilation interpretation.

The Appendices complement the discussion here with extended results.  In particular, App.~\ref{sec:roi} includes further justification for the choice of ROI and App.~\ref{sec:siginj} summarizes signal injection and recovery tests. Extended results, including limits for different annihilation channels and DM profiles, as well as other variations of the astrophysical templates, are provided in App.~\ref{sec:extended}.  \vspace{0.05in}

\section{Acknowledgements}

We thank Keith~Bechtol, Ilias~Cholis, Tim~Cohen, Dan~Hooper, Tim~Linden, Nick~Rodd, Ben~Safdi and Christoph~Weniger for useful conversations. We thank Nick~Rodd for producing the data used in this study, originally created for the analysis in~\cite{Cohen:2016uyg}. In addition to aforementioned software packages, this research made use of the \texttt{Astropy}~\cite{2013A&A...558A..33A}, \texttt{IPython}~\cite{PER-GRA:2007} and \texttt{Minuit}~\cite{James:1975dr} packages. LJC is supported by a Paul and Daisy Soros Fellowship and an NSF Graduate Research Fellowship. ML is supported by the DOE under contract DESC0007968, the Alfred P. Sloan Foundation and the Cottrell Scholar Program through the Research Corporation for Science
Advancement. 

\bibliographystyle{apsrev}
\bibliography{fermi_smoothgal}

\begin{thebibliography}{79}
\expandafter\ifx\csname natexlab\endcsname\relax\def\natexlab#1{#1}\fi
\expandafter\ifx\csname bibnamefont\endcsname\relax
  \def\bibnamefont#1{#1}\fi
\expandafter\ifx\csname bibfnamefont\endcsname\relax
  \def\bibfnamefont#1{#1}\fi
\expandafter\ifx\csname citenamefont\endcsname\relax
  \def\citenamefont#1{#1}\fi
\expandafter\ifx\csname url\endcsname\relax
  \def\url#1{\texttt{#1}}\fi
\expandafter\ifx\csname urlprefix\endcsname\relax\def\urlprefix{URL }\fi
\providecommand{\bibinfo}[2]{#2}
\providecommand{\eprint}[2][]{\url{#2}}

\bibitem[{\citenamefont{Atwood et~al.}(2009)}]{Atwood:2009ez}
\bibinfo{author}{\bibfnamefont{W.~B.} \bibnamefont{Atwood}}
  \bibnamefont{et~al.} (\bibinfo{collaboration}{Fermi-LAT}),
  \bibinfo{journal}{Astrophys. J.} \textbf{\bibinfo{volume}{697}},
  \bibinfo{pages}{1071} (\bibinfo{year}{2009}), \eprint{0902.1089}.

\bibitem[{\citenamefont{Ackermann
  et~al.}(2014{\natexlab{a}})}]{Ackermann:2013yva}
\bibinfo{author}{\bibfnamefont{M.}~\bibnamefont{Ackermann}}
  \bibnamefont{et~al.} (\bibinfo{collaboration}{Fermi-LAT}),
  \bibinfo{journal}{Phys. Rev.} \textbf{\bibinfo{volume}{D89}},
  \bibinfo{pages}{042001} (\bibinfo{year}{2014}{\natexlab{a}}),
  \eprint{1310.0828}.

\bibitem[{\citenamefont{Geringer-Sameth
  et~al.}(2015{\natexlab{a}})\citenamefont{Geringer-Sameth, Koushiappas, and
  Walker}}]{Geringer-Sameth:2014qqa}
\bibinfo{author}{\bibfnamefont{A.}~\bibnamefont{Geringer-Sameth}},
  \bibinfo{author}{\bibfnamefont{S.~M.} \bibnamefont{Koushiappas}},
  \bibnamefont{and} \bibinfo{author}{\bibfnamefont{M.~G.}
  \bibnamefont{Walker}}, \bibinfo{journal}{Phys. Rev.}
  \textbf{\bibinfo{volume}{D91}}, \bibinfo{pages}{083535}
  (\bibinfo{year}{2015}{\natexlab{a}}), \eprint{1410.2242}.

\bibitem[{\citenamefont{Ackermann
  et~al.}(2015{\natexlab{a}})}]{Ackermann:2015zua}
\bibinfo{author}{\bibfnamefont{M.}~\bibnamefont{Ackermann}}
  \bibnamefont{et~al.} (\bibinfo{collaboration}{Fermi-LAT}),
  \bibinfo{journal}{Phys. Rev. Lett.} \textbf{\bibinfo{volume}{115}},
  \bibinfo{pages}{231301} (\bibinfo{year}{2015}{\natexlab{a}}),
  \eprint{1503.02641}.

\bibitem[{\citenamefont{Albert et~al.}(2017)}]{Fermi-LAT:2016uux}
\bibinfo{author}{\bibfnamefont{A.}~\bibnamefont{Albert}} \bibnamefont{et~al.}
  (\bibinfo{collaboration}{DES, Fermi-LAT}), \bibinfo{journal}{Astrophys. J.}
  \textbf{\bibinfo{volume}{834}}, \bibinfo{pages}{110} (\bibinfo{year}{2017}),
  \eprint{1611.03184}.

\bibitem[{\citenamefont{Lisanti
  et~al.}(2017{\natexlab{a}})\citenamefont{Lisanti, Mishra-Sharma, Rodd, Safdi,
  and Wechsler}}]{Lisanti:2017qoz}
\bibinfo{author}{\bibfnamefont{M.}~\bibnamefont{Lisanti}},
  \bibinfo{author}{\bibfnamefont{S.}~\bibnamefont{Mishra-Sharma}},
  \bibinfo{author}{\bibfnamefont{N.~L.} \bibnamefont{Rodd}},
  \bibinfo{author}{\bibfnamefont{B.~R.} \bibnamefont{Safdi}}, \bibnamefont{and}
  \bibinfo{author}{\bibfnamefont{R.~H.} \bibnamefont{Wechsler}}
  (\bibinfo{year}{2017}{\natexlab{a}}), \eprint{1709.00416}.

\bibitem[{\citenamefont{Lisanti
  et~al.}(2017{\natexlab{b}})\citenamefont{Lisanti, Mishra-Sharma, Rodd, and
  Safdi}}]{Lisanti:2017qlb}
\bibinfo{author}{\bibfnamefont{M.}~\bibnamefont{Lisanti}},
  \bibinfo{author}{\bibfnamefont{S.}~\bibnamefont{Mishra-Sharma}},
  \bibinfo{author}{\bibfnamefont{N.~L.} \bibnamefont{Rodd}}, \bibnamefont{and}
  \bibinfo{author}{\bibfnamefont{B.~R.} \bibnamefont{Safdi}}
  (\bibinfo{year}{2017}{\natexlab{b}}), \eprint{1708.09385}.

\bibitem[{\citenamefont{Ackermann
  et~al.}(2015{\natexlab{b}})}]{Ackermann:2014usa}
\bibinfo{author}{\bibfnamefont{M.}~\bibnamefont{Ackermann}}
  \bibnamefont{et~al.} (\bibinfo{collaboration}{Fermi-LAT}),
  \bibinfo{journal}{Astrophys. J.} \textbf{\bibinfo{volume}{799}},
  \bibinfo{pages}{86} (\bibinfo{year}{2015}{\natexlab{b}}), \eprint{1410.3696}.

\bibitem[{\citenamefont{Goodenough and Hooper}(2009)}]{Goodenough:2009gk}
\bibinfo{author}{\bibfnamefont{L.}~\bibnamefont{Goodenough}} \bibnamefont{and}
  \bibinfo{author}{\bibfnamefont{D.}~\bibnamefont{Hooper}}
  (\bibinfo{year}{2009}), \eprint{0910.2998}.

\bibitem[{\citenamefont{Hooper and Goodenough}(2011)}]{Hooper:2010mq}
\bibinfo{author}{\bibfnamefont{D.}~\bibnamefont{Hooper}} \bibnamefont{and}
  \bibinfo{author}{\bibfnamefont{L.}~\bibnamefont{Goodenough}},
  \bibinfo{journal}{Phys. Lett.} \textbf{\bibinfo{volume}{B697}},
  \bibinfo{pages}{412} (\bibinfo{year}{2011}), \eprint{1010.2752}.

\bibitem[{\citenamefont{Boyarsky et~al.}(2011)\citenamefont{Boyarsky, Malyshev,
  and Ruchayskiy}}]{Boyarsky:2010dr}
\bibinfo{author}{\bibfnamefont{A.}~\bibnamefont{Boyarsky}},
  \bibinfo{author}{\bibfnamefont{D.}~\bibnamefont{Malyshev}}, \bibnamefont{and}
  \bibinfo{author}{\bibfnamefont{O.}~\bibnamefont{Ruchayskiy}},
  \bibinfo{journal}{Phys. Lett.} \textbf{\bibinfo{volume}{B705}},
  \bibinfo{pages}{165} (\bibinfo{year}{2011}), \eprint{1012.5839}.

\bibitem[{\citenamefont{Hooper and Linden}(2011)}]{Hooper:2011ti}
\bibinfo{author}{\bibfnamefont{D.}~\bibnamefont{Hooper}} \bibnamefont{and}
  \bibinfo{author}{\bibfnamefont{T.}~\bibnamefont{Linden}},
  \bibinfo{journal}{Phys. Rev.} \textbf{\bibinfo{volume}{D84}},
  \bibinfo{pages}{123005} (\bibinfo{year}{2011}), \eprint{1110.0006}.

\bibitem[{\citenamefont{Abazajian and Kaplinghat}(2012)}]{Abazajian:2012pn}
\bibinfo{author}{\bibfnamefont{K.~N.} \bibnamefont{Abazajian}}
  \bibnamefont{and}
  \bibinfo{author}{\bibfnamefont{M.}~\bibnamefont{Kaplinghat}},
  \bibinfo{journal}{Phys. Rev.} \textbf{\bibinfo{volume}{D86}},
  \bibinfo{pages}{083511} (\bibinfo{year}{2012}), \bibinfo{note}{[Erratum:
  Phys. Rev.D87,129902(2013)]}, \eprint{1207.6047}.

\bibitem[{\citenamefont{Hooper and Slatyer}(2013)}]{Hooper:2013rwa}
\bibinfo{author}{\bibfnamefont{D.}~\bibnamefont{Hooper}} \bibnamefont{and}
  \bibinfo{author}{\bibfnamefont{T.~R.} \bibnamefont{Slatyer}},
  \bibinfo{journal}{Phys. Dark Univ.} \textbf{\bibinfo{volume}{2}},
  \bibinfo{pages}{118} (\bibinfo{year}{2013}), \eprint{1302.6589}.

\bibitem[{\citenamefont{Gordon and Macias}(2013)}]{Gordon:2013vta}
\bibinfo{author}{\bibfnamefont{C.}~\bibnamefont{Gordon}} \bibnamefont{and}
  \bibinfo{author}{\bibfnamefont{O.}~\bibnamefont{Macias}},
  \bibinfo{journal}{Phys. Rev.} \textbf{\bibinfo{volume}{D88}},
  \bibinfo{pages}{083521} (\bibinfo{year}{2013}), \bibinfo{note}{[Erratum:
  Phys. Rev.D89,no.4,049901(2014)]}, \eprint{1306.5725}.

\bibitem[{\citenamefont{Huang et~al.}(2013)\citenamefont{Huang, Urbano, and
  Xue}}]{Huang:2013pda}
\bibinfo{author}{\bibfnamefont{W.-C.} \bibnamefont{Huang}},
  \bibinfo{author}{\bibfnamefont{A.}~\bibnamefont{Urbano}}, \bibnamefont{and}
  \bibinfo{author}{\bibfnamefont{W.}~\bibnamefont{Xue}} (\bibinfo{year}{2013}),
  \eprint{1307.6862}.

\bibitem[{\citenamefont{Macias and Gordon}(2014)}]{Macias:2013vya}
\bibinfo{author}{\bibfnamefont{O.}~\bibnamefont{Macias}} \bibnamefont{and}
  \bibinfo{author}{\bibfnamefont{C.}~\bibnamefont{Gordon}},
  \bibinfo{journal}{Phys. Rev.} \textbf{\bibinfo{volume}{D89}},
  \bibinfo{pages}{063515} (\bibinfo{year}{2014}), \eprint{1312.6671}.

\bibitem[{\citenamefont{Abazajian et~al.}(2014)\citenamefont{Abazajian, Canac,
  Horiuchi, and Kaplinghat}}]{Abazajian:2014fta}
\bibinfo{author}{\bibfnamefont{K.~N.} \bibnamefont{Abazajian}},
  \bibinfo{author}{\bibfnamefont{N.}~\bibnamefont{Canac}},
  \bibinfo{author}{\bibfnamefont{S.}~\bibnamefont{Horiuchi}}, \bibnamefont{and}
  \bibinfo{author}{\bibfnamefont{M.}~\bibnamefont{Kaplinghat}},
  \bibinfo{journal}{Phys. Rev.} \textbf{\bibinfo{volume}{D90}},
  \bibinfo{pages}{023526} (\bibinfo{year}{2014}), \eprint{1402.4090}.

\bibitem[{\citenamefont{Daylan et~al.}(2016)\citenamefont{Daylan, Finkbeiner,
  Hooper, Linden, Portillo, Rodd, and Slatyer}}]{Daylan:2014rsa}
\bibinfo{author}{\bibfnamefont{T.}~\bibnamefont{Daylan}},
  \bibinfo{author}{\bibfnamefont{D.~P.} \bibnamefont{Finkbeiner}},
  \bibinfo{author}{\bibfnamefont{D.}~\bibnamefont{Hooper}},
  \bibinfo{author}{\bibfnamefont{T.}~\bibnamefont{Linden}},
  \bibinfo{author}{\bibfnamefont{S.~K.~N.} \bibnamefont{Portillo}},
  \bibinfo{author}{\bibfnamefont{N.~L.} \bibnamefont{Rodd}}, \bibnamefont{and}
  \bibinfo{author}{\bibfnamefont{T.~R.} \bibnamefont{Slatyer}},
  \bibinfo{journal}{Phys. Dark Univ.} \textbf{\bibinfo{volume}{12}},
  \bibinfo{pages}{1} (\bibinfo{year}{2016}), \eprint{1402.6703}.

\bibitem[{\citenamefont{Zhou et~al.}(2015)\citenamefont{Zhou, Liang, Huang, Li,
  Fan, Feng, and Chang}}]{Zhou:2014lva}
\bibinfo{author}{\bibfnamefont{B.}~\bibnamefont{Zhou}},
  \bibinfo{author}{\bibfnamefont{Y.-F.} \bibnamefont{Liang}},
  \bibinfo{author}{\bibfnamefont{X.}~\bibnamefont{Huang}},
  \bibinfo{author}{\bibfnamefont{X.}~\bibnamefont{Li}},
  \bibinfo{author}{\bibfnamefont{Y.-Z.} \bibnamefont{Fan}},
  \bibinfo{author}{\bibfnamefont{L.}~\bibnamefont{Feng}}, \bibnamefont{and}
  \bibinfo{author}{\bibfnamefont{J.}~\bibnamefont{Chang}},
  \bibinfo{journal}{Phys. Rev.} \textbf{\bibinfo{volume}{D91}},
  \bibinfo{pages}{123010} (\bibinfo{year}{2015}), \eprint{1406.6948}.

\bibitem[{\citenamefont{Calore et~al.}(2015{\natexlab{a}})\citenamefont{Calore,
  Cholis, and Weniger}}]{Calore:2014xka}
\bibinfo{author}{\bibfnamefont{F.}~\bibnamefont{Calore}},
  \bibinfo{author}{\bibfnamefont{I.}~\bibnamefont{Cholis}}, \bibnamefont{and}
  \bibinfo{author}{\bibfnamefont{C.}~\bibnamefont{Weniger}},
  \bibinfo{journal}{JCAP} \textbf{\bibinfo{volume}{1503}}, \bibinfo{pages}{038}
  (\bibinfo{year}{2015}{\natexlab{a}}), \eprint{1409.0042}.

\bibitem[{\citenamefont{Abazajian et~al.}(2015)\citenamefont{Abazajian, Canac,
  Horiuchi, Kaplinghat, and Kwa}}]{Abazajian:2014hsa}
\bibinfo{author}{\bibfnamefont{K.~N.} \bibnamefont{Abazajian}},
  \bibinfo{author}{\bibfnamefont{N.}~\bibnamefont{Canac}},
  \bibinfo{author}{\bibfnamefont{S.}~\bibnamefont{Horiuchi}},
  \bibinfo{author}{\bibfnamefont{M.}~\bibnamefont{Kaplinghat}},
  \bibnamefont{and} \bibinfo{author}{\bibfnamefont{A.}~\bibnamefont{Kwa}},
  \bibinfo{journal}{JCAP} \textbf{\bibinfo{volume}{1507}}, \bibinfo{pages}{013}
  (\bibinfo{year}{2015}), \eprint{1410.6168}.

\bibitem[{\citenamefont{Ajello et~al.}(2016)}]{TheFermi-LAT:2015kwa}
\bibinfo{author}{\bibfnamefont{M.}~\bibnamefont{Ajello}} \bibnamefont{et~al.}
  (\bibinfo{collaboration}{Fermi-LAT}), \bibinfo{journal}{Astrophys. J.}
  \textbf{\bibinfo{volume}{819}}, \bibinfo{pages}{44} (\bibinfo{year}{2016}),
  \eprint{1511.02938}.

\bibitem[{\citenamefont{Karwin et~al.}(2017)\citenamefont{Karwin, Murgia, Tait,
  Porter, and Tanedo}}]{Karwin:2016tsw}
\bibinfo{author}{\bibfnamefont{C.}~\bibnamefont{Karwin}},
  \bibinfo{author}{\bibfnamefont{S.}~\bibnamefont{Murgia}},
  \bibinfo{author}{\bibfnamefont{T.~M.~P.} \bibnamefont{Tait}},
  \bibinfo{author}{\bibfnamefont{T.~A.} \bibnamefont{Porter}},
  \bibnamefont{and} \bibinfo{author}{\bibfnamefont{P.}~\bibnamefont{Tanedo}},
  \bibinfo{journal}{Phys. Rev.} \textbf{\bibinfo{volume}{D95}},
  \bibinfo{pages}{103005} (\bibinfo{year}{2017}), \eprint{1612.05687}.

\bibitem[{\citenamefont{Lee et~al.}(2016)\citenamefont{Lee, Lisanti, Safdi,
  Slatyer, and Xue}}]{Lee:2015fea}
\bibinfo{author}{\bibfnamefont{S.~K.} \bibnamefont{Lee}},
  \bibinfo{author}{\bibfnamefont{M.}~\bibnamefont{Lisanti}},
  \bibinfo{author}{\bibfnamefont{B.~R.} \bibnamefont{Safdi}},
  \bibinfo{author}{\bibfnamefont{T.~R.} \bibnamefont{Slatyer}},
  \bibnamefont{and} \bibinfo{author}{\bibfnamefont{W.}~\bibnamefont{Xue}},
  \bibinfo{journal}{Phys. Rev. Lett.} \textbf{\bibinfo{volume}{116}},
  \bibinfo{pages}{051103} (\bibinfo{year}{2016}), \eprint{1506.05124}.

\bibitem[{\citenamefont{Bartels et~al.}(2016)\citenamefont{Bartels,
  Krishnamurthy, and Weniger}}]{Bartels:2015aea}
\bibinfo{author}{\bibfnamefont{R.}~\bibnamefont{Bartels}},
  \bibinfo{author}{\bibfnamefont{S.}~\bibnamefont{Krishnamurthy}},
  \bibnamefont{and} \bibinfo{author}{\bibfnamefont{C.}~\bibnamefont{Weniger}},
  \bibinfo{journal}{Phys. Rev. Lett.} \textbf{\bibinfo{volume}{116}},
  \bibinfo{pages}{051102} (\bibinfo{year}{2016}), \eprint{1506.05104}.

\bibitem[{\citenamefont{Macias et~al.}(2016)\citenamefont{Macias, Gordon,
  Crocker, Coleman, Paterson, Horiuchi, and Pohl}}]{Macias:2016nev}
\bibinfo{author}{\bibfnamefont{O.}~\bibnamefont{Macias}},
  \bibinfo{author}{\bibfnamefont{C.}~\bibnamefont{Gordon}},
  \bibinfo{author}{\bibfnamefont{R.~M.} \bibnamefont{Crocker}},
  \bibinfo{author}{\bibfnamefont{B.}~\bibnamefont{Coleman}},
  \bibinfo{author}{\bibfnamefont{D.}~\bibnamefont{Paterson}},
  \bibinfo{author}{\bibfnamefont{S.}~\bibnamefont{Horiuchi}}, \bibnamefont{and}
  \bibinfo{author}{\bibfnamefont{M.}~\bibnamefont{Pohl}}
  (\bibinfo{year}{2016}), \eprint{1611.06644}.

\bibitem[{\citenamefont{Bartels et~al.}(2017)\citenamefont{Bartels, Storm,
  Weniger, and Calore}}]{Bartels:2017vsx}
\bibinfo{author}{\bibfnamefont{R.}~\bibnamefont{Bartels}},
  \bibinfo{author}{\bibfnamefont{E.}~\bibnamefont{Storm}},
  \bibinfo{author}{\bibfnamefont{C.}~\bibnamefont{Weniger}}, \bibnamefont{and}
  \bibinfo{author}{\bibfnamefont{F.}~\bibnamefont{Calore}}
  (\bibinfo{year}{2017}), \eprint{1711.04778}.

\bibitem[{\citenamefont{Keeley et~al.}(2017)\citenamefont{Keeley, Abazajian,
  Kwa, Rodd, and Safdi}}]{Keeley:2017fbz}
\bibinfo{author}{\bibfnamefont{R.}~\bibnamefont{Keeley}},
  \bibinfo{author}{\bibfnamefont{K.}~\bibnamefont{Abazajian}},
  \bibinfo{author}{\bibfnamefont{A.}~\bibnamefont{Kwa}},
  \bibinfo{author}{\bibfnamefont{N.}~\bibnamefont{Rodd}}, \bibnamefont{and}
  \bibinfo{author}{\bibfnamefont{B.}~\bibnamefont{Safdi}}
  (\bibinfo{year}{2017}), \eprint{1710.03215}.

\bibitem[{\citenamefont{Sanders et~al.}(2016)\citenamefont{Sanders, Evans,
  Geringer-Sameth, and Dehnen}}]{Sanders:2016eie}
\bibinfo{author}{\bibfnamefont{J.~L.} \bibnamefont{Sanders}},
  \bibinfo{author}{\bibfnamefont{N.~W.} \bibnamefont{Evans}},
  \bibinfo{author}{\bibfnamefont{A.}~\bibnamefont{Geringer-Sameth}},
  \bibnamefont{and} \bibinfo{author}{\bibfnamefont{W.}~\bibnamefont{Dehnen}},
  \bibinfo{journal}{Phys. Rev.} \textbf{\bibinfo{volume}{D94}},
  \bibinfo{pages}{063521} (\bibinfo{year}{2016}), \eprint{1604.05493}.

\bibitem[{\citenamefont{{Bonnivard} et~al.}(2016)\citenamefont{{Bonnivard},
  {Maurin}, and {Walker}}}]{2016MNRAS.462..223B}
\bibinfo{author}{\bibfnamefont{V.}~\bibnamefont{{Bonnivard}}},
  \bibinfo{author}{\bibfnamefont{D.}~\bibnamefont{{Maurin}}}, \bibnamefont{and}
  \bibinfo{author}{\bibfnamefont{M.~G.} \bibnamefont{{Walker}}},
  \bibinfo{journal}{MNRAS} \textbf{\bibinfo{volume}{462}}, \bibinfo{pages}{223}
  (\bibinfo{year}{2016}), \eprint{1506.08209}.

\bibitem[{\citenamefont{Geringer-Sameth
  et~al.}(2015{\natexlab{b}})\citenamefont{Geringer-Sameth, Koushiappas, and
  Walker}}]{Geringer-Sameth:2014yza}
\bibinfo{author}{\bibfnamefont{A.}~\bibnamefont{Geringer-Sameth}},
  \bibinfo{author}{\bibfnamefont{S.~M.} \bibnamefont{Koushiappas}},
  \bibnamefont{and} \bibinfo{author}{\bibfnamefont{M.}~\bibnamefont{Walker}},
  \bibinfo{journal}{Astrophys. J.} \textbf{\bibinfo{volume}{801}},
  \bibinfo{pages}{74} (\bibinfo{year}{2015}{\natexlab{b}}), \eprint{1408.0002}.

\bibitem[{\citenamefont{Abazajian and Keeley}(2016)}]{Abazajian:2015raa}
\bibinfo{author}{\bibfnamefont{K.~N.} \bibnamefont{Abazajian}}
  \bibnamefont{and} \bibinfo{author}{\bibfnamefont{R.~E.}
  \bibnamefont{Keeley}}, \bibinfo{journal}{Phys. Rev.}
  \textbf{\bibinfo{volume}{D93}}, \bibinfo{pages}{083514}
  (\bibinfo{year}{2016}), \eprint{1510.06424}.

\bibitem[{\citenamefont{Papucci and Strumia}(2010)}]{Papucci:2009gd}
\bibinfo{author}{\bibfnamefont{M.}~\bibnamefont{Papucci}} \bibnamefont{and}
  \bibinfo{author}{\bibfnamefont{A.}~\bibnamefont{Strumia}},
  \bibinfo{journal}{JCAP} \textbf{\bibinfo{volume}{1003}}, \bibinfo{pages}{014}
  (\bibinfo{year}{2010}), \eprint{0912.0742}.

\bibitem[{\citenamefont{Cirelli et~al.}(2010)\citenamefont{Cirelli, Panci, and
  Serpico}}]{Cirelli:2009dv}
\bibinfo{author}{\bibfnamefont{M.}~\bibnamefont{Cirelli}},
  \bibinfo{author}{\bibfnamefont{P.}~\bibnamefont{Panci}}, \bibnamefont{and}
  \bibinfo{author}{\bibfnamefont{P.~D.} \bibnamefont{Serpico}},
  \bibinfo{journal}{Nucl. Phys.} \textbf{\bibinfo{volume}{B840}},
  \bibinfo{pages}{284} (\bibinfo{year}{2010}), \eprint{0912.0663}.

\bibitem[{\citenamefont{Baxter and Dodelson}(2011)}]{Baxter:2011rc}
\bibinfo{author}{\bibfnamefont{E.~J.} \bibnamefont{Baxter}} \bibnamefont{and}
  \bibinfo{author}{\bibfnamefont{S.}~\bibnamefont{Dodelson}},
  \bibinfo{journal}{Phys. Rev.} \textbf{\bibinfo{volume}{D83}},
  \bibinfo{pages}{123516} (\bibinfo{year}{2011}), \eprint{1103.5779}.

\bibitem[{\citenamefont{Malyshev et~al.}(2011)\citenamefont{Malyshev, Bovy, and
  Cholis}}]{Malyshev:2010zzc}
\bibinfo{author}{\bibfnamefont{D.}~\bibnamefont{Malyshev}},
  \bibinfo{author}{\bibfnamefont{J.}~\bibnamefont{Bovy}}, \bibnamefont{and}
  \bibinfo{author}{\bibfnamefont{I.}~\bibnamefont{Cholis}},
  \bibinfo{journal}{Phys. Rev.} \textbf{\bibinfo{volume}{D84}},
  \bibinfo{pages}{023013} (\bibinfo{year}{2011}), \eprint{1007.4556}.

\bibitem[{\citenamefont{Ackermann
  et~al.}(2012{\natexlab{a}})}]{Ackermann:2012rg}
\bibinfo{author}{\bibfnamefont{M.}~\bibnamefont{Ackermann}}
  \bibnamefont{et~al.} (\bibinfo{collaboration}{Fermi-LAT}),
  \bibinfo{journal}{Astrophys. J.} \textbf{\bibinfo{volume}{761}},
  \bibinfo{pages}{91} (\bibinfo{year}{2012}{\natexlab{a}}), \eprint{1205.6474}.

\bibitem[{\citenamefont{Zechlin et~al.}(2017)\citenamefont{Zechlin, Manconi,
  and Donato}}]{Zechlin:2017uzo}
\bibinfo{author}{\bibfnamefont{H.~S.} \bibnamefont{Zechlin}},
  \bibinfo{author}{\bibfnamefont{S.}~\bibnamefont{Manconi}}, \bibnamefont{and}
  \bibinfo{author}{\bibfnamefont{F.}~\bibnamefont{Donato}}
  (\bibinfo{year}{2017}), \eprint{1710.01506}.

\bibitem[{\citenamefont{Huang et~al.}(2016)\citenamefont{Huang, En{\ss}lin, and
  Selig}}]{Huang:2015rlu}
\bibinfo{author}{\bibfnamefont{X.}~\bibnamefont{Huang}},
  \bibinfo{author}{\bibfnamefont{T.}~\bibnamefont{En{\ss}lin}},
  \bibnamefont{and} \bibinfo{author}{\bibfnamefont{M.}~\bibnamefont{Selig}},
  \bibinfo{journal}{JCAP} \textbf{\bibinfo{volume}{1604}}, \bibinfo{pages}{030}
  (\bibinfo{year}{2016}), \eprint{1511.02621}.

\bibitem[{\citenamefont{Gorski et~al.}(2005)\citenamefont{Gorski, Hivon,
  Banday, Wandelt, Hansen, Reinecke, and Bartelman}}]{Gorski:2004by}
\bibinfo{author}{\bibfnamefont{K.~M.} \bibnamefont{Gorski}},
  \bibinfo{author}{\bibfnamefont{E.}~\bibnamefont{Hivon}},
  \bibinfo{author}{\bibfnamefont{A.~J.} \bibnamefont{Banday}},
  \bibinfo{author}{\bibfnamefont{B.~D.} \bibnamefont{Wandelt}},
  \bibinfo{author}{\bibfnamefont{F.~K.} \bibnamefont{Hansen}},
  \bibinfo{author}{\bibfnamefont{M.}~\bibnamefont{Reinecke}}, \bibnamefont{and}
  \bibinfo{author}{\bibfnamefont{M.}~\bibnamefont{Bartelman}},
  \bibinfo{journal}{Astrophys. J.} \textbf{\bibinfo{volume}{622}},
  \bibinfo{pages}{759} (\bibinfo{year}{2005}), \eprint{astro-ph/0409513}.

\bibitem[{\citenamefont{Gaggero
  et~al.}(2015{\natexlab{a}})\citenamefont{Gaggero, Taoso, Urbano, Valli, and
  Ullio}}]{Gaggero:2015nsa}
\bibinfo{author}{\bibfnamefont{D.}~\bibnamefont{Gaggero}},
  \bibinfo{author}{\bibfnamefont{M.}~\bibnamefont{Taoso}},
  \bibinfo{author}{\bibfnamefont{A.}~\bibnamefont{Urbano}},
  \bibinfo{author}{\bibfnamefont{M.}~\bibnamefont{Valli}}, \bibnamefont{and}
  \bibinfo{author}{\bibfnamefont{P.}~\bibnamefont{Ullio}},
  \bibinfo{journal}{JCAP} \textbf{\bibinfo{volume}{1512}}, \bibinfo{pages}{056}
  (\bibinfo{year}{2015}{\natexlab{a}}), \eprint{1507.06129}.

\bibitem[{\citenamefont{Werner et~al.}(2015)\citenamefont{Werner, Kissmann,
  Strong, and Reimer}}]{Werner:2014sya}
\bibinfo{author}{\bibfnamefont{M.}~\bibnamefont{Werner}},
  \bibinfo{author}{\bibfnamefont{R.}~\bibnamefont{Kissmann}},
  \bibinfo{author}{\bibfnamefont{A.~W.} \bibnamefont{Strong}},
  \bibnamefont{and} \bibinfo{author}{\bibfnamefont{O.}~\bibnamefont{Reimer}},
  \bibinfo{journal}{Astropart. Phys.} \textbf{\bibinfo{volume}{64}},
  \bibinfo{pages}{18} (\bibinfo{year}{2015}), \eprint{1410.5266}.

\bibitem[{\citenamefont{Kissmann}(2014)}]{Kissmann:2014sia}
\bibinfo{author}{\bibfnamefont{R.}~\bibnamefont{Kissmann}},
  \bibinfo{journal}{Astropart. Phys.} \textbf{\bibinfo{volume}{55}},
  \bibinfo{pages}{37} (\bibinfo{year}{2014}), \eprint{1401.4035}.

\bibitem[{\citenamefont{Gaggero
  et~al.}(2015{\natexlab{b}})\citenamefont{Gaggero, Urbano, Valli, and
  Ullio}}]{Gaggero:2014xla}
\bibinfo{author}{\bibfnamefont{D.}~\bibnamefont{Gaggero}},
  \bibinfo{author}{\bibfnamefont{A.}~\bibnamefont{Urbano}},
  \bibinfo{author}{\bibfnamefont{M.}~\bibnamefont{Valli}}, \bibnamefont{and}
  \bibinfo{author}{\bibfnamefont{P.}~\bibnamefont{Ullio}},
  \bibinfo{journal}{Phys. Rev.} \textbf{\bibinfo{volume}{D91}},
  \bibinfo{pages}{083012} (\bibinfo{year}{2015}{\natexlab{b}}),
  \eprint{1411.7623}.

\bibitem[{\citenamefont{Evoli et~al.}(2017)\citenamefont{Evoli, Gaggero,
  Vittino, Di~Bernardo, Di~Mauro, Ligorini, Ullio, and Grasso}}]{Evoli:2016xgn}
\bibinfo{author}{\bibfnamefont{C.}~\bibnamefont{Evoli}},
  \bibinfo{author}{\bibfnamefont{D.}~\bibnamefont{Gaggero}},
  \bibinfo{author}{\bibfnamefont{A.}~\bibnamefont{Vittino}},
  \bibinfo{author}{\bibfnamefont{G.}~\bibnamefont{Di~Bernardo}},
  \bibinfo{author}{\bibfnamefont{M.}~\bibnamefont{Di~Mauro}},
  \bibinfo{author}{\bibfnamefont{A.}~\bibnamefont{Ligorini}},
  \bibinfo{author}{\bibfnamefont{P.}~\bibnamefont{Ullio}}, \bibnamefont{and}
  \bibinfo{author}{\bibfnamefont{D.}~\bibnamefont{Grasso}},
  \bibinfo{journal}{JCAP} \textbf{\bibinfo{volume}{1702}}, \bibinfo{pages}{015}
  (\bibinfo{year}{2017}), \eprint{1607.07886}.

\bibitem[{\citenamefont{Storm et~al.}(2017)\citenamefont{Storm, Weniger, and
  Calore}}]{Storm:2017arh}
\bibinfo{author}{\bibfnamefont{E.}~\bibnamefont{Storm}},
  \bibinfo{author}{\bibfnamefont{C.}~\bibnamefont{Weniger}}, \bibnamefont{and}
  \bibinfo{author}{\bibfnamefont{F.}~\bibnamefont{Calore}},
  \bibinfo{journal}{JCAP} \textbf{\bibinfo{volume}{1708}}, \bibinfo{pages}{022}
  (\bibinfo{year}{2017}), \eprint{1705.04065}.

\bibitem[{\citenamefont{Selig et~al.}(2015)\citenamefont{Selig, Vacca,
  Oppermann, and En{\ss}lin}}]{Selig:2014qqa}
\bibinfo{author}{\bibfnamefont{M.}~\bibnamefont{Selig}},
  \bibinfo{author}{\bibfnamefont{V.}~\bibnamefont{Vacca}},
  \bibinfo{author}{\bibfnamefont{N.}~\bibnamefont{Oppermann}},
  \bibnamefont{and} \bibinfo{author}{\bibfnamefont{T.~A.}
  \bibnamefont{En{\ss}lin}}, \bibinfo{journal}{Astron. Astrophys.}
  \textbf{\bibinfo{volume}{581}}, \bibinfo{pages}{A126} (\bibinfo{year}{2015}),
  \eprint{1410.4562}.

\bibitem[{\citenamefont{Su et~al.}(2010)\citenamefont{Su, Slatyer, and
  Finkbeiner}}]{Su:2010qj}
\bibinfo{author}{\bibfnamefont{M.}~\bibnamefont{Su}},
  \bibinfo{author}{\bibfnamefont{T.~R.} \bibnamefont{Slatyer}},
  \bibnamefont{and} \bibinfo{author}{\bibfnamefont{D.~P.}
  \bibnamefont{Finkbeiner}}, \bibinfo{journal}{Astrophys. J.}
  \textbf{\bibinfo{volume}{724}}, \bibinfo{pages}{1044} (\bibinfo{year}{2010}),
  \eprint{1005.5480}.

\bibitem[{\citenamefont{Acero et~al.}(2015)}]{Acero:2015hja}
\bibinfo{author}{\bibfnamefont{F.}~\bibnamefont{Acero}} \bibnamefont{et~al.}
  (\bibinfo{collaboration}{Fermi-LAT}), \bibinfo{journal}{Astrophys. J. Suppl.}
  \textbf{\bibinfo{volume}{218}}, \bibinfo{pages}{23} (\bibinfo{year}{2015}),
  \eprint{1501.02003}.

\bibitem[{\citenamefont{Navarro et~al.}(1997)\citenamefont{Navarro, Frenk, and
  White}}]{Navarro:1996gj}
\bibinfo{author}{\bibfnamefont{J.~F.} \bibnamefont{Navarro}},
  \bibinfo{author}{\bibfnamefont{C.~S.} \bibnamefont{Frenk}}, \bibnamefont{and}
  \bibinfo{author}{\bibfnamefont{S.~D.~M.} \bibnamefont{White}},
  \bibinfo{journal}{Astrophys. J.} \textbf{\bibinfo{volume}{490}},
  \bibinfo{pages}{493} (\bibinfo{year}{1997}), \eprint{astro-ph/9611107}.

\bibitem[{\citenamefont{{McKee} et~al.}(2015)\citenamefont{{McKee},
  {Parravano}, and {Hollenbach}}}]{2015ApJ...814...13M}
\bibinfo{author}{\bibfnamefont{C.~F.} \bibnamefont{{McKee}}},
  \bibinfo{author}{\bibfnamefont{A.}~\bibnamefont{{Parravano}}},
  \bibnamefont{and} \bibinfo{author}{\bibfnamefont{D.~J.}
  \bibnamefont{{Hollenbach}}}, \bibinfo{journal}{\apj}
  \textbf{\bibinfo{volume}{814}}, \bibinfo{eid}{13} (\bibinfo{year}{2015}),
  \eprint{1509.05334}.

\bibitem[{\citenamefont{Sivertsson et~al.}(2017)\citenamefont{Sivertsson,
  Silverwood, Read, Bertone, and Steger}}]{Sivertsson:2017rkp}
\bibinfo{author}{\bibfnamefont{S.}~\bibnamefont{Sivertsson}},
  \bibinfo{author}{\bibfnamefont{H.}~\bibnamefont{Silverwood}},
  \bibinfo{author}{\bibfnamefont{J.~I.} \bibnamefont{Read}},
  \bibinfo{author}{\bibfnamefont{G.}~\bibnamefont{Bertone}}, \bibnamefont{and}
  \bibinfo{author}{\bibfnamefont{P.}~\bibnamefont{Steger}},
  \bibinfo{journal}{Submitted to: Mon. Not. Roy. Astron. Soc.}
  (\bibinfo{year}{2017}), \eprint{1708.07836}.

\bibitem[{\citenamefont{Read}(2014)}]{Read:2014qva}
\bibinfo{author}{\bibfnamefont{J.~I.} \bibnamefont{Read}}, \bibinfo{journal}{J.
  Phys.} \textbf{\bibinfo{volume}{G41}}, \bibinfo{pages}{063101}
  (\bibinfo{year}{2014}), \eprint{1404.1938}.

\bibitem[{\citenamefont{Strong et~al.}(2000)\citenamefont{Strong, Moskalenko,
  and Reimer}}]{Strong:1998fr}
\bibinfo{author}{\bibfnamefont{A.~W.} \bibnamefont{Strong}},
  \bibinfo{author}{\bibfnamefont{I.~V.} \bibnamefont{Moskalenko}},
  \bibnamefont{and} \bibinfo{author}{\bibfnamefont{O.}~\bibnamefont{Reimer}},
  \bibinfo{journal}{Astrophys. J.} \textbf{\bibinfo{volume}{537}},
  \bibinfo{pages}{763} (\bibinfo{year}{2000}), \bibinfo{note}{[Erratum:
  Astrophys. J.541,1109(2000)]}, \eprint{astro-ph/9811296}.

\bibitem[{\citenamefont{Acero et~al.}(2016)}]{Acero:2016qlg}
\bibinfo{author}{\bibfnamefont{F.}~\bibnamefont{Acero}} \bibnamefont{et~al.}
  (\bibinfo{collaboration}{Fermi-LAT}), \bibinfo{journal}{Astrophys. J. Suppl.}
  \textbf{\bibinfo{volume}{223}}, \bibinfo{pages}{26} (\bibinfo{year}{2016}),
  \eprint{1602.07246}.

\bibitem[{\citenamefont{{Large} et~al.}(1962)\citenamefont{{Large}, {Quigley},
  and {Haslam}}}]{1962MNRAS.124..405L}
\bibinfo{author}{\bibfnamefont{M.~I.} \bibnamefont{{Large}}},
  \bibinfo{author}{\bibfnamefont{M.~J.~S.} \bibnamefont{{Quigley}}},
  \bibnamefont{and} \bibinfo{author}{\bibfnamefont{C.~G.~T.}
  \bibnamefont{{Haslam}}}, \bibinfo{journal}{MNRAS}
  \textbf{\bibinfo{volume}{124}}, \bibinfo{pages}{405} (\bibinfo{year}{1962}).

\bibitem[{\citenamefont{{Haslam} et~al.}(1981)\citenamefont{{Haslam}, {Klein},
  {Salter}, {Stoffel}, {Wilson}, {Cleary}, {Cooke}, and
  {Thomasson}}}]{1981A&A...100..209H}
\bibinfo{author}{\bibfnamefont{C.~G.~T.} \bibnamefont{{Haslam}}},
  \bibinfo{author}{\bibfnamefont{U.}~\bibnamefont{{Klein}}},
  \bibinfo{author}{\bibfnamefont{C.~J.} \bibnamefont{{Salter}}},
  \bibinfo{author}{\bibfnamefont{H.}~\bibnamefont{{Stoffel}}},
  \bibinfo{author}{\bibfnamefont{W.~E.} \bibnamefont{{Wilson}}},
  \bibinfo{author}{\bibfnamefont{M.~N.} \bibnamefont{{Cleary}}},
  \bibinfo{author}{\bibfnamefont{D.~J.} \bibnamefont{{Cooke}}},
  \bibnamefont{and}
  \bibinfo{author}{\bibfnamefont{P.}~\bibnamefont{{Thomasson}}},
  \bibinfo{journal}{Astron. Astrophys.} \textbf{\bibinfo{volume}{100}},
  \bibinfo{pages}{209} (\bibinfo{year}{1981}).

\bibitem[{\citenamefont{{Casandjian} et~al.}(2009)\citenamefont{{Casandjian},
  {Grenier}, and {for the Fermi Large Area Telescope
  Collaboration}}}]{2009arXiv0912.3478C}
\bibinfo{author}{\bibfnamefont{J.-M.} \bibnamefont{{Casandjian}}},
  \bibinfo{author}{\bibfnamefont{I.}~\bibnamefont{{Grenier}}},
  \bibnamefont{and} \bibinfo{author}{\bibnamefont{{for the Fermi Large Area
  Telescope Collaboration}}}, \bibinfo{journal}{ArXiv e-prints}
  (\bibinfo{year}{2009}), \eprint{0912.3478}.

\bibitem[{\citenamefont{Ackermann
  et~al.}(2012{\natexlab{b}})}]{Ackermann:2012uf}
\bibinfo{author}{\bibfnamefont{M.}~\bibnamefont{Ackermann}}
  \bibnamefont{et~al.} (\bibinfo{collaboration}{Fermi-LAT}),
  \bibinfo{journal}{Phys. Rev.} \textbf{\bibinfo{volume}{D85}},
  \bibinfo{pages}{083007} (\bibinfo{year}{2012}{\natexlab{b}}),
  \eprint{1202.2856}.

\bibitem[{\citenamefont{Ackermann
  et~al.}(2014{\natexlab{b}})}]{Fermi-LAT:2014sfa}
\bibinfo{author}{\bibfnamefont{M.}~\bibnamefont{Ackermann}}
  \bibnamefont{et~al.} (\bibinfo{collaboration}{Fermi-LAT}),
  \bibinfo{journal}{Astrophys. J.} \textbf{\bibinfo{volume}{793}},
  \bibinfo{pages}{64} (\bibinfo{year}{2014}{\natexlab{b}}), \eprint{1407.7905}.

\bibitem[{\citenamefont{Rolke et~al.}(2005)\citenamefont{Rolke, Lopez, and
  Conrad}}]{Rolke:2004mj}
\bibinfo{author}{\bibfnamefont{W.~A.} \bibnamefont{Rolke}},
  \bibinfo{author}{\bibfnamefont{A.~M.} \bibnamefont{Lopez}}, \bibnamefont{and}
  \bibinfo{author}{\bibfnamefont{J.}~\bibnamefont{Conrad}},
  \bibinfo{journal}{Nucl. Instrum. Meth.} \textbf{\bibinfo{volume}{A551}},
  \bibinfo{pages}{493} (\bibinfo{year}{2005}), \eprint{physics/0403059}.

\bibitem[{\citenamefont{Cirelli et~al.}(2011)\citenamefont{Cirelli, Corcella,
  Hektor, Hutsi, Kadastik, Panci, Raidal, Sala, and Strumia}}]{Cirelli:2010xx}
\bibinfo{author}{\bibfnamefont{M.}~\bibnamefont{Cirelli}},
  \bibinfo{author}{\bibfnamefont{G.}~\bibnamefont{Corcella}},
  \bibinfo{author}{\bibfnamefont{A.}~\bibnamefont{Hektor}},
  \bibinfo{author}{\bibfnamefont{G.}~\bibnamefont{Hutsi}},
  \bibinfo{author}{\bibfnamefont{M.}~\bibnamefont{Kadastik}},
  \bibinfo{author}{\bibfnamefont{P.}~\bibnamefont{Panci}},
  \bibinfo{author}{\bibfnamefont{M.}~\bibnamefont{Raidal}},
  \bibinfo{author}{\bibfnamefont{F.}~\bibnamefont{Sala}}, \bibnamefont{and}
  \bibinfo{author}{\bibfnamefont{A.}~\bibnamefont{Strumia}},
  \bibinfo{journal}{JCAP} \textbf{\bibinfo{volume}{1103}}, \bibinfo{pages}{051}
  (\bibinfo{year}{2011}), \bibinfo{note}{[Erratum: JCAP1210,E01(2012)]},
  \eprint{1012.4515}.

\bibitem[{\citenamefont{Mishra-Sharma et~al.}(2017)\citenamefont{Mishra-Sharma,
  Rodd, and Safdi}}]{Mishra-Sharma:2016gis}
\bibinfo{author}{\bibfnamefont{S.}~\bibnamefont{Mishra-Sharma}},
  \bibinfo{author}{\bibfnamefont{N.~L.} \bibnamefont{Rodd}}, \bibnamefont{and}
  \bibinfo{author}{\bibfnamefont{B.~R.} \bibnamefont{Safdi}},
  \bibinfo{journal}{Astron. J.} \textbf{\bibinfo{volume}{153}},
  \bibinfo{pages}{253} (\bibinfo{year}{2017}), \eprint{1612.03173}.

\bibitem[{\citenamefont{Byrd et~al.}(1995)\citenamefont{Byrd, Lu, Nocedal, and
  Zhu}}]{citeulike:10176711}
\bibinfo{author}{\bibfnamefont{R.~H.} \bibnamefont{Byrd}},
  \bibinfo{author}{\bibfnamefont{P.}~\bibnamefont{Lu}},
  \bibinfo{author}{\bibfnamefont{J.}~\bibnamefont{Nocedal}}, \bibnamefont{and}
  \bibinfo{author}{\bibfnamefont{C.}~\bibnamefont{Zhu}}, \bibinfo{journal}{SIAM
  J. Sci. Comput.} \textbf{\bibinfo{volume}{16}}, \bibinfo{pages}{1190}
  (\bibinfo{year}{1995}), ISSN \bibinfo{issn}{1064-8275},
  \urlprefix\url{http://dx.doi.org/10.1137/0916069}.

\bibitem[{\citenamefont{Jones et~al.}(2001--)\citenamefont{Jones, Oliphant,
  Peterson et~al.}}]{scipycite}
\bibinfo{author}{\bibfnamefont{E.}~\bibnamefont{Jones}},
  \bibinfo{author}{\bibfnamefont{T.}~\bibnamefont{Oliphant}},
  \bibinfo{author}{\bibfnamefont{P.}~\bibnamefont{Peterson}},
  \bibnamefont{et~al.}, \emph{\bibinfo{title}{{SciPy}: Open source scientific
  tools for {Python}}} (\bibinfo{year}{2001--}),
  \urlprefix\url{http://www.scipy.org/}.

\bibitem[{\citenamefont{Steigman et~al.}(2012)\citenamefont{Steigman, Dasgupta,
  and Beacom}}]{Steigman:2012nb}
\bibinfo{author}{\bibfnamefont{G.}~\bibnamefont{Steigman}},
  \bibinfo{author}{\bibfnamefont{B.}~\bibnamefont{Dasgupta}}, \bibnamefont{and}
  \bibinfo{author}{\bibfnamefont{J.~F.} \bibnamefont{Beacom}},
  \bibinfo{journal}{Phys. Rev.} \textbf{\bibinfo{volume}{D86}},
  \bibinfo{pages}{023506} (\bibinfo{year}{2012}), \eprint{1204.3622}.

\bibitem[{\citenamefont{{Ackermann} et~al.}(2012)\citenamefont{{Ackermann},
  {Ajello}, {Atwood}, {Baldini}, {Ballet}, {Barbiellini}, {Bastieri},
  {Bechtol}, {Bellazzini}, {Berenji} et~al.}}]{2012ApJ...750....3A}
\bibinfo{author}{\bibfnamefont{M.}~\bibnamefont{{Ackermann}}},
  \bibinfo{author}{\bibfnamefont{M.}~\bibnamefont{{Ajello}}},
  \bibinfo{author}{\bibfnamefont{W.~B.} \bibnamefont{{Atwood}}},
  \bibinfo{author}{\bibfnamefont{L.}~\bibnamefont{{Baldini}}},
  \bibinfo{author}{\bibfnamefont{J.}~\bibnamefont{{Ballet}}},
  \bibinfo{author}{\bibfnamefont{G.}~\bibnamefont{{Barbiellini}}},
  \bibinfo{author}{\bibfnamefont{D.}~\bibnamefont{{Bastieri}}},
  \bibinfo{author}{\bibfnamefont{K.}~\bibnamefont{{Bechtol}}},
  \bibinfo{author}{\bibfnamefont{R.}~\bibnamefont{{Bellazzini}}},
  \bibinfo{author}{\bibfnamefont{B.}~\bibnamefont{{Berenji}}},
  \bibnamefont{et~al.}, \bibinfo{journal}{\apj} \textbf{\bibinfo{volume}{750}},
  \bibinfo{eid}{3} (\bibinfo{year}{2012}), \eprint{1202.4039}.

\bibitem[{\citenamefont{Cohen et~al.}(2017)\citenamefont{Cohen, Murase, Rodd,
  Safdi, and Soreq}}]{Cohen:2016uyg}
\bibinfo{author}{\bibfnamefont{T.}~\bibnamefont{Cohen}},
  \bibinfo{author}{\bibfnamefont{K.}~\bibnamefont{Murase}},
  \bibinfo{author}{\bibfnamefont{N.~L.} \bibnamefont{Rodd}},
  \bibinfo{author}{\bibfnamefont{B.~R.} \bibnamefont{Safdi}}, \bibnamefont{and}
  \bibinfo{author}{\bibfnamefont{Y.}~\bibnamefont{Soreq}},
  \bibinfo{journal}{Phys. Rev. Lett.} \textbf{\bibinfo{volume}{119}},
  \bibinfo{pages}{021102} (\bibinfo{year}{2017}), \eprint{1612.05638}.

\bibitem[{\citenamefont{{Astropy Collaboration}
  et~al.}(2013)\citenamefont{{Astropy Collaboration}, {Robitaille}, {Tollerud},
  {Greenfield}, {Droettboom}, {Bray}, {Aldcroft}, {Davis}, {Ginsburg},
  {Price-Whelan} et~al.}}]{2013A&A...558A..33A}
\bibinfo{author}{\bibnamefont{{Astropy Collaboration}}},
  \bibinfo{author}{\bibfnamefont{T.~P.} \bibnamefont{{Robitaille}}},
  \bibinfo{author}{\bibfnamefont{E.~J.} \bibnamefont{{Tollerud}}},
  \bibinfo{author}{\bibfnamefont{P.}~\bibnamefont{{Greenfield}}},
  \bibinfo{author}{\bibfnamefont{M.}~\bibnamefont{{Droettboom}}},
  \bibinfo{author}{\bibfnamefont{E.}~\bibnamefont{{Bray}}},
  \bibinfo{author}{\bibfnamefont{T.}~\bibnamefont{{Aldcroft}}},
  \bibinfo{author}{\bibfnamefont{M.}~\bibnamefont{{Davis}}},
  \bibinfo{author}{\bibfnamefont{A.}~\bibnamefont{{Ginsburg}}},
  \bibinfo{author}{\bibfnamefont{A.~M.} \bibnamefont{{Price-Whelan}}},
  \bibnamefont{et~al.}, \bibinfo{journal}{AAP} \textbf{\bibinfo{volume}{558}},
  \bibinfo{eid}{A33} (\bibinfo{year}{2013}), \eprint{1307.6212}.

\bibitem[{\citenamefont{P\'erez and Granger}(2007)}]{PER-GRA:2007}
\bibinfo{author}{\bibfnamefont{F.}~\bibnamefont{P\'erez}} \bibnamefont{and}
  \bibinfo{author}{\bibfnamefont{B.~E.} \bibnamefont{Granger}},
  \bibinfo{journal}{Computing in Science and Engineering}
  \textbf{\bibinfo{volume}{9}}, \bibinfo{pages}{21} (\bibinfo{year}{2007}),
  ISSN \bibinfo{issn}{1521-9615}, \urlprefix\url{http://ipython.org}.

\bibitem[{\citenamefont{James and Roos}(1975)}]{James:1975dr}
\bibinfo{author}{\bibfnamefont{F.}~\bibnamefont{James}} \bibnamefont{and}
  \bibinfo{author}{\bibfnamefont{M.}~\bibnamefont{Roos}},
  \bibinfo{journal}{Comput. Phys. Commun.} \textbf{\bibinfo{volume}{10}},
  \bibinfo{pages}{343} (\bibinfo{year}{1975}).

\bibitem[{\citenamefont{Cowan et~al.}(2011)\citenamefont{Cowan, Cranmer, Gross,
  and Vitells}}]{Cowan:2010js}
\bibinfo{author}{\bibfnamefont{G.}~\bibnamefont{Cowan}},
  \bibinfo{author}{\bibfnamefont{K.}~\bibnamefont{Cranmer}},
  \bibinfo{author}{\bibfnamefont{E.}~\bibnamefont{Gross}}, \bibnamefont{and}
  \bibinfo{author}{\bibfnamefont{O.}~\bibnamefont{Vitells}},
  \bibinfo{journal}{Eur. Phys. J.} \textbf{\bibinfo{volume}{C71}},
  \bibinfo{pages}{1554} (\bibinfo{year}{2011}), \bibinfo{note}{[Erratum: Eur.
  Phys. J.C73,2501(2013)]}, \eprint{1007.1727}.

\bibitem[{\citenamefont{{Einasto}}(1965)}]{1965TrAlm...5...87E}
\bibinfo{author}{\bibfnamefont{J.}~\bibnamefont{{Einasto}}},
  \bibinfo{journal}{Trudy Astrofizicheskogo Instituta Alma-Ata}
  \textbf{\bibinfo{volume}{5}}, \bibinfo{pages}{87} (\bibinfo{year}{1965}).

\bibitem[{\citenamefont{Navarro et~al.}(2010)\citenamefont{Navarro, Ludlow,
  Springel, Wang, Vogelsberger, White, Jenkins, Frenk, and
  Helmi}}]{Navarro:2008kc}
\bibinfo{author}{\bibfnamefont{J.~F.} \bibnamefont{Navarro}},
  \bibinfo{author}{\bibfnamefont{A.}~\bibnamefont{Ludlow}},
  \bibinfo{author}{\bibfnamefont{V.}~\bibnamefont{Springel}},
  \bibinfo{author}{\bibfnamefont{J.}~\bibnamefont{Wang}},
  \bibinfo{author}{\bibfnamefont{M.}~\bibnamefont{Vogelsberger}},
  \bibinfo{author}{\bibfnamefont{S.~D.~M.} \bibnamefont{White}},
  \bibinfo{author}{\bibfnamefont{A.}~\bibnamefont{Jenkins}},
  \bibinfo{author}{\bibfnamefont{C.~S.} \bibnamefont{Frenk}}, \bibnamefont{and}
  \bibinfo{author}{\bibfnamefont{A.}~\bibnamefont{Helmi}},
  \bibinfo{journal}{Mon. Not. Roy. Astron. Soc.}
  \textbf{\bibinfo{volume}{402}}, \bibinfo{pages}{21} (\bibinfo{year}{2010}),
  \eprint{0810.1522}.

\bibitem[{\citenamefont{Burkert}(1996)}]{Burkert:1995yz}
\bibinfo{author}{\bibfnamefont{A.}~\bibnamefont{Burkert}},
  \bibinfo{journal}{IAU Symp.} \textbf{\bibinfo{volume}{171}},
  \bibinfo{pages}{175} (\bibinfo{year}{1996}), \bibinfo{note}{[Astrophys.
  J.447,L25(1995)]}, \eprint{astro-ph/9504041}.

\bibitem[{\citenamefont{Maccio' et~al.}(2012)\citenamefont{Maccio', Stinson,
  Brook, Wadsley, Couchman, Shen, Gibson, and Quinn}}]{Maccio:2011ryn}
\bibinfo{author}{\bibfnamefont{A.~V.} \bibnamefont{Maccio'}},
  \bibinfo{author}{\bibfnamefont{G.}~\bibnamefont{Stinson}},
  \bibinfo{author}{\bibfnamefont{C.~B.} \bibnamefont{Brook}},
  \bibinfo{author}{\bibfnamefont{J.}~\bibnamefont{Wadsley}},
  \bibinfo{author}{\bibfnamefont{H.~M.~P.} \bibnamefont{Couchman}},
  \bibinfo{author}{\bibfnamefont{S.}~\bibnamefont{Shen}},
  \bibinfo{author}{\bibfnamefont{B.~K.} \bibnamefont{Gibson}},
  \bibnamefont{and} \bibinfo{author}{\bibfnamefont{T.}~\bibnamefont{Quinn}},
  \bibinfo{journal}{Astrophys. J.} \textbf{\bibinfo{volume}{744}},
  \bibinfo{pages}{L9} (\bibinfo{year}{2012}), \eprint{1111.5620}.

\bibitem[{\citenamefont{Di~Cintio et~al.}(2014)\citenamefont{Di~Cintio, Brook,
  Macci{\`o}, Stinson, Knebe, Dutton, and Wadsley}}]{DiCintio:2013qxa}
\bibinfo{author}{\bibfnamefont{A.}~\bibnamefont{Di~Cintio}},
  \bibinfo{author}{\bibfnamefont{C.~B.} \bibnamefont{Brook}},
  \bibinfo{author}{\bibfnamefont{A.~V.} \bibnamefont{Macci{\`o}}},
  \bibinfo{author}{\bibfnamefont{G.~S.} \bibnamefont{Stinson}},
  \bibinfo{author}{\bibfnamefont{A.}~\bibnamefont{Knebe}},
  \bibinfo{author}{\bibfnamefont{A.~A.} \bibnamefont{Dutton}},
  \bibnamefont{and} \bibinfo{author}{\bibfnamefont{J.}~\bibnamefont{Wadsley}},
  \bibinfo{journal}{Mon. Not. Roy. Astron. Soc.}
  \textbf{\bibinfo{volume}{437}}, \bibinfo{pages}{415} (\bibinfo{year}{2014}),
  \eprint{1306.0898}.

\bibitem[{\citenamefont{Calore et~al.}(2015{\natexlab{b}})\citenamefont{Calore,
  Cholis, McCabe, and Weniger}}]{Calore:2014nla}
\bibinfo{author}{\bibfnamefont{F.}~\bibnamefont{Calore}},
  \bibinfo{author}{\bibfnamefont{I.}~\bibnamefont{Cholis}},
  \bibinfo{author}{\bibfnamefont{C.}~\bibnamefont{McCabe}}, \bibnamefont{and}
  \bibinfo{author}{\bibfnamefont{C.}~\bibnamefont{Weniger}},
  \bibinfo{journal}{Phys. Rev.} \textbf{\bibinfo{volume}{D91}},
  \bibinfo{pages}{063003} (\bibinfo{year}{2015}{\natexlab{b}}),
  \eprint{1411.4647}.

\end{thebibliography}

\appendix

\section{The Region-of-Interest}
\label{sec:roi}

\begin{figure*}[htbp]
\centering
\includegraphics[width=\textwidth]{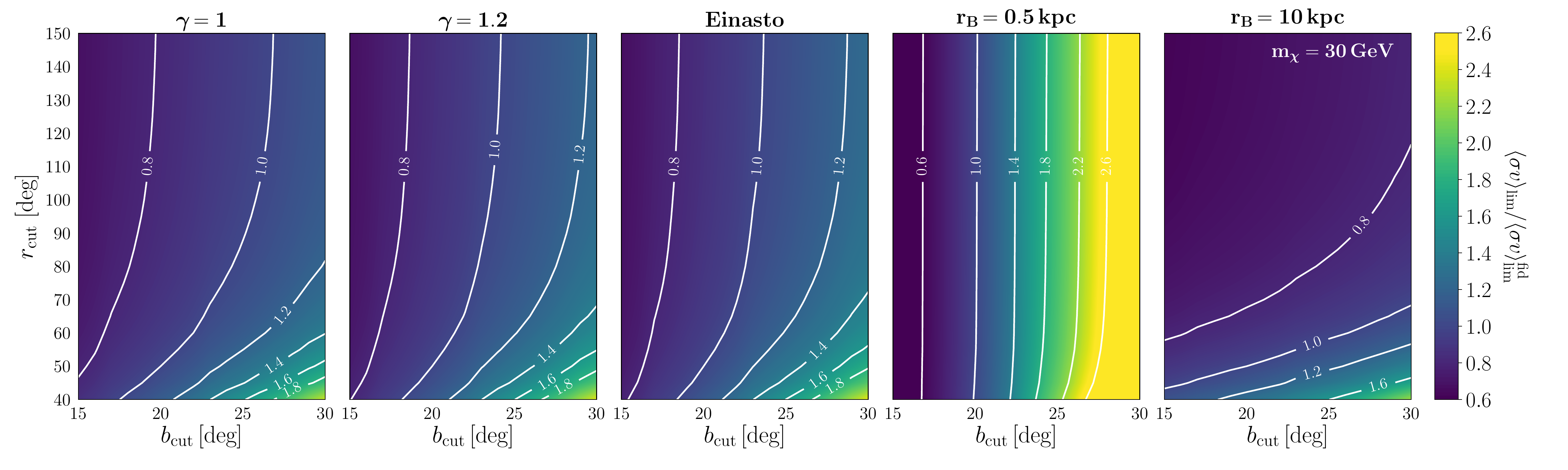} \hspace{4mm}
\caption{Sensitivity projections for a 30~GeV dark matter particle annihilating to $b\bar{b}$ for different regions of interest, which are defined by latitude ($|b| > b_\text{cut}$) and radial ($r < r_\text{cut}$) cuts.   The projected limit, $\langle \sigma v \rangle_\text{lim}$, is compared to the limit for the fiducial region, $\langle \sigma v \rangle_\text{lim}^\text{fid}$, which corresponds to $|b| > 20^\circ$ and $r < 50^\circ$.  The contours indicate the ratio of these two cross sections.  The projections are provided for different dark matter density profiles:  (left to right) generalized NFW with inner-slope $\gamma = 1, 1.2$, Einasto, and Burkert with a $r_B=0.5$ and $10$~kpc core.}
\label{fig:asimov}
\end{figure*}

The fiducial analysis presented in this paper uses a region-of-interest (ROI) defined by the annulus $|b| > 20^\circ$ and $r < 50^\circ$.  To motivate this choice, we analyze Asimov datasets~\cite{Cowan:2010js}, which can be used to determine the median asymptotic behavior of the test statistic under the assumption that the foregrounds are perfectly modeled, while varying over different choices of the ROI. The Asimov dataset in this case corresponds to the sum of astrophysical templates best-fit to the data in each ROI.  Note that the \texttt{p6v11} template was not divided into independent radial slices in the Asimov study.

As a concrete example, we consider the case of a 30~GeV DM particle annihilating to $b\bar{b}$, although results for other DM masses are largely unchanged. We vary over latitude ($|b| > b_\text{cut}$) and radial ($r < r_\text{cut}$) cuts spanning $b_\text{cut}=\{15^\circ,16^\circ,\ldots,30^\circ\}$ and $r_\text{cut} = \{40^\circ,45^\circ,\ldots,150^\circ\}$. Figure~\ref{fig:asimov} demonstrates how the projected cross section limit, $\langle \sigma v \rangle_\text{lim}$, compares to that for the fiducial ROI, $\langle \sigma v \rangle^\text{fid}_\text{lim}$, as a function of $b_\text{cut}$ and $r_\text{cut}$. We consider the generalized NFW profile as in Eq.~\ref{eq:NFW} with scale radius $r_s = 17$~kpc, local density $\rho(r_\odot) = 0.4$~GeV$\,$cm$^{-3}$, and inner slope $\gamma = 1$ and $1.2$ (first and second panel from left, respectively).  In general, we see that the projected sensitivity strengthens for smaller $b_\text{cut}$ and larger $r_\text{cut}$, as expected.  This dependence weakens for steeper profiles because the dark matter (DM) density is concentrated towards the Galactic Center. We note that the Asimov projections assume perfect knowledge of the astrophysical components, and as such disregard potential degeneracies between a DM signal and astrophysical templates,  which are likely to be important in an analysis on data.

For comparison, we also consider several other DM density profiles, each normalized to $\rho(r_\odot) = 0.4$~GeV$\,$cm$^{-3}$.  The middle panel of Fig.~\ref{fig:asimov} shows the results for the Einasto profile~\cite{1965TrAlm...5...87E}:
\begin{equation}
\rho_{\rm Einasto}(r) = \rho_0\exp\left[-\frac{2}{\alpha}\left(\left(\frac{r}{r_E}\right)^\alpha-1\right)\right] \, ,
\label{eq:Einasto}
\end{equation}
with $\alpha=0.17$ and $r_E = 15.14$~kpc~\cite{Navarro:2008kc}.  The final two panels in Fig.~\ref{fig:asimov} show the results for a cored Burkert profile~\cite{Burkert:1995yz}:
\begin{equation}
\rho_{\rm Burkert}(r) = \frac{\rho_0}{(1+r/r_B)[1+(r/r_B)^2]}\,,
\label{eq:Burkert}
\end{equation}
where $r_B$ is the analog of the NFW  scale radius and sets the size of the core.  For illustration, we consider $r_B = 0.5$ and $10$~kpc, which roughly spans the range of allowed possibilities---see \emph{e.g.}~\cite{Maccio:2011ryn,DiCintio:2013qxa}. While the Einasto contours look very similar to those for NFW with $\gamma = 1.2$, the Burkert results are quite different.  For the smaller core, there is only very mild dependence on $r_\text{cut}$ and the projected signal strength decreases with larger $b_\text{cut}$.  In contrast, the signal is strengthened with decreased latitude and increased radial cuts for the case where $r_B = 10$~kpc because the DM distribution is less concentrated towards the Galactic Center.

\section{Signal Injection and Recovery}
\label{sec:siginj}

A vital consistency check involves ensuring that the limit-setting procedure would not exclude a DM signal if one were present in the data. We perform a variety of tests to confirm that we can set a robust limit while recovering the properties of a DM signal. We perform these checks on both Monte Carlo simulations as well as on the data itself.\\

\noindent
\textbf{Signal injection on Monte Carlo.}  We create Monte Carlo simulations of the gamma-ray sky by summing the astrophysical templates best-fit on data, adding the signal from a DM particle annihilating to $b\bar b$ in the smooth Galactic halo, and Poisson fluctuating the final map. We create 50 Monte Carlo realizations of the sky map and pass each through the analysis pipeline.  This procedure is repeated for different DM masses and cross sections to study the resulting limit and the test statistic associated with the extracted signal.  

\begin{figure*}[htbp]
\centering
\includegraphics[width=0.48\textwidth]{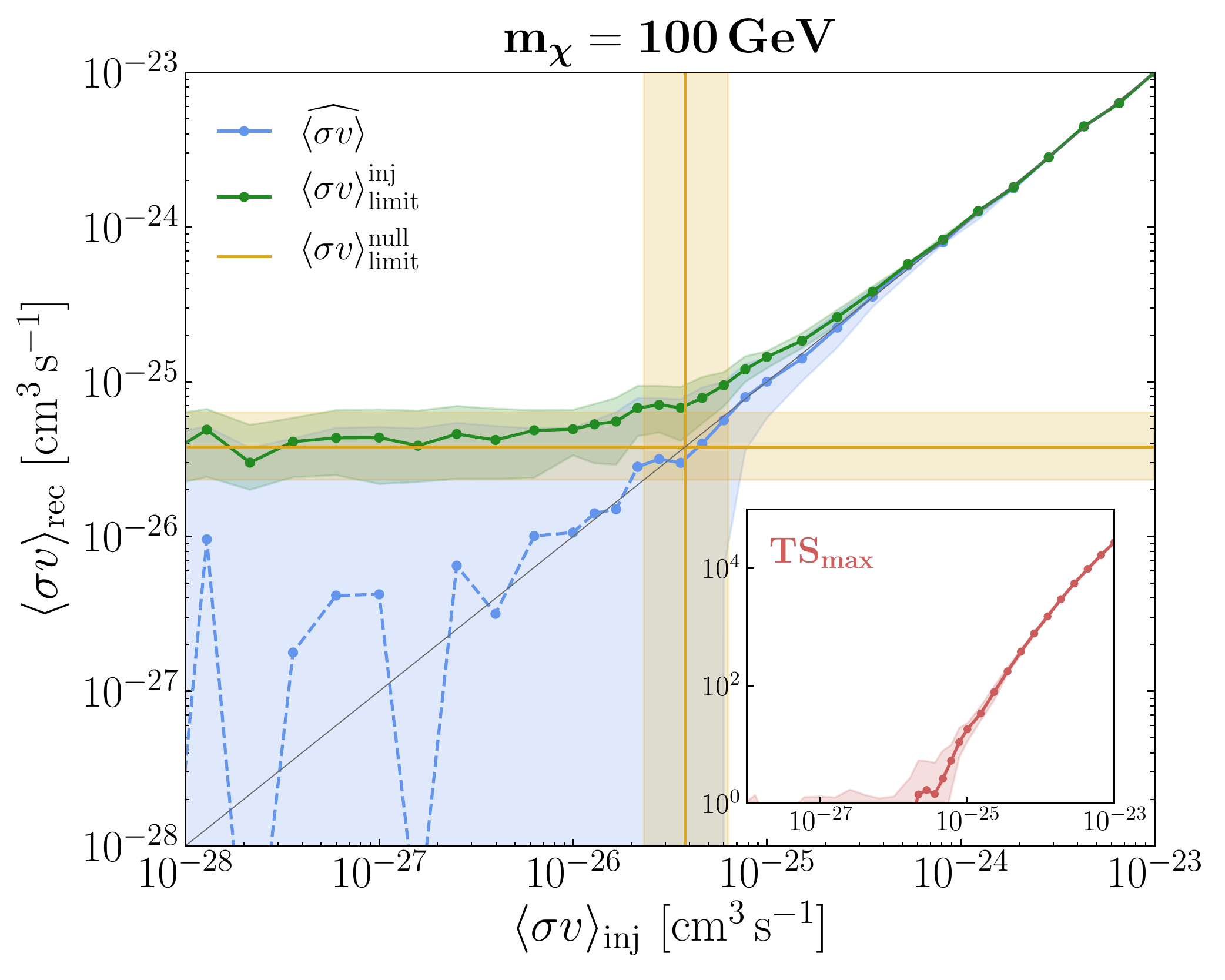} \hspace{4mm}
\includegraphics[width=0.48\textwidth]{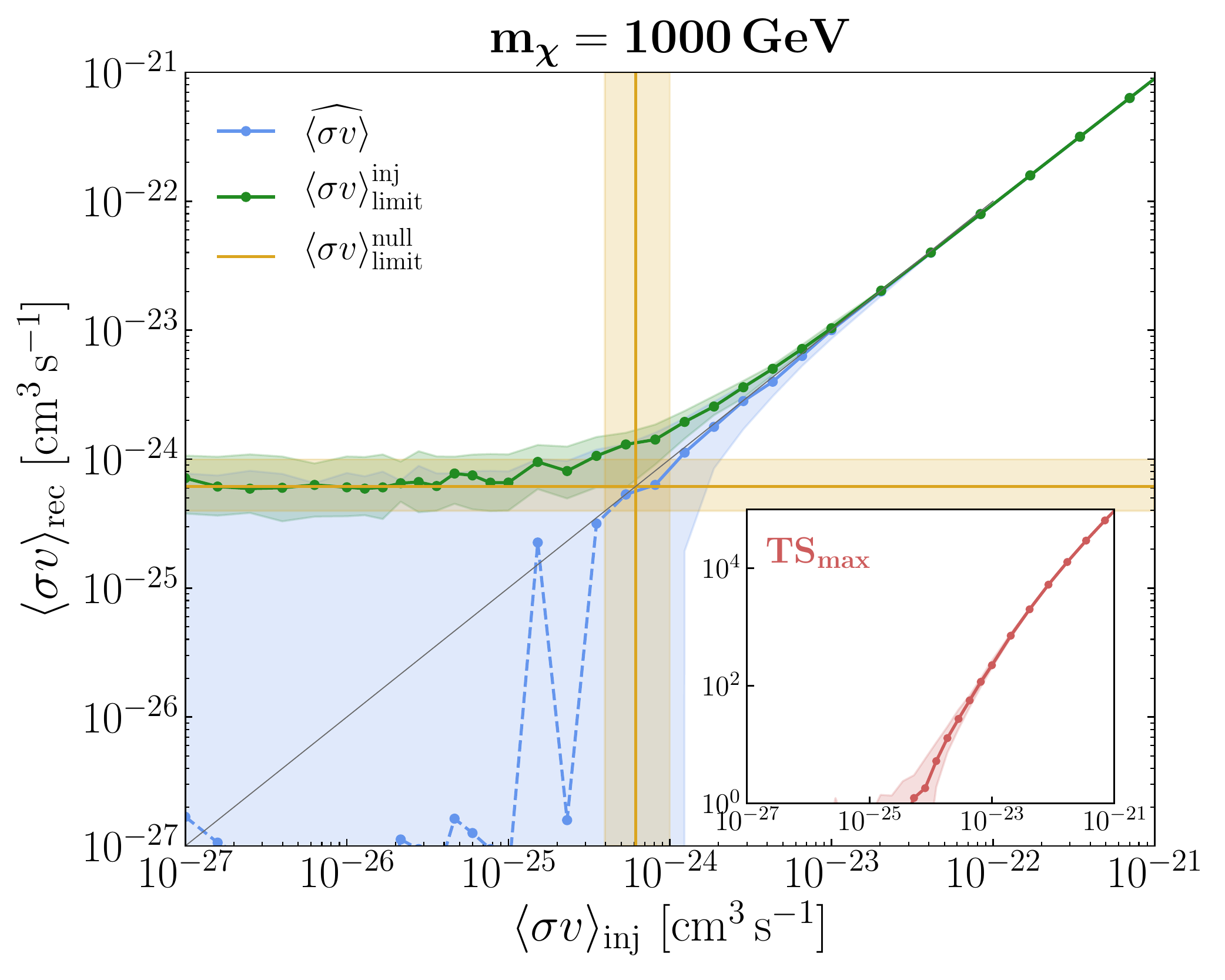}
\caption{Signal injection tests on Monte Carlo simulations for a 100 (left) and 1000 (right)~GeV DM particle annihilating to $b\bar{b}$.  In each panel, the gold line corresponds to the limit $\langle \sigma v \rangle_\text{limit}^\text{null}$ obtained when no signal is injected into the simulated data.  The green line corresponds to the median cross section limit, $\langle \sigma v \rangle_\text{limit}^\text{inj}$, that is recovered for a given injected cross section $\langle \sigma v \rangle_\text{inj}$, when 
$\text{TS} = -2.71$. The green band shows the corresponding 68\% containment. The blue line corresponds to the median recovered cross section $\widehat{\langle \sigma v\rangle}$ that is associated with the maximum test statistic $\text{TS}_\text{max}$ (plotted in the inset), and is shown as dashed in the regime where $\text{TS}_\text{max} < 1$.  The blue band spans extremal values of the 68\% containment of cross sections associated with TS$_\text{max}-1$.   For each injected signal point, we create 50 realizations of simulated sky maps.}
\label{fig:injsigmc}
\end{figure*}

Figure~\ref{fig:injsigmc} summarizes the results of the signal injection tests for $m_\chi = 100$ and $1000$~GeV in the left and right panel, respectively.  In each panel, the gold bands indicate the recovered limit, $\langle \sigma v \rangle_\text{limit}^\text{null}$, when no signal is injected into the simulated sky map.  The green band shows the middle 68\% containment of the cross section, $\langle \sigma v \rangle_\text{limit}^\text{inj}$, that is recovered when $\text{TS} = -2.71$ in the presence of an injected signal with cross section $\langle \sigma v \rangle_\text{inj}$.  If the statistical procedure is robust, the green band should lie above the diagonal line (saying that the limit set would be consistent with an injected signal) and should asymptotically approach the gold band for small signal cross sections, as is indeed the case for both masses included here. 

The blue line shows the recovered cross section that is associated with the maximum test statistic, TS$_\text{max}$:
\begin{eqnarray}
{\rm TS}_\text{max} \equiv 2 \, \big[&& \log \mathcal{L}(d | \mathcal{M}, \widehat{\langle\sigma v\rangle}, m_\chi )  \nonumber \\
&& - \log \mathcal{L}(d | \mathcal{M}, \langle\sigma v\rangle = 0, m_\chi ) \big]\,, 
\label{TSmax}
\end{eqnarray}
where $\widehat{\langle\sigma v\rangle}$ is the cross section that maximizes the likelihood. In the regime where $\text{TS}_\text{max} < 1$, this is shown as a dashed line. The blue band corresponds to the range of cross sections above and below $\widehat{\langle\sigma v\rangle}$ associated with TS$_\text{max}-1$, spanning the extremal values of the middle 68\% containment in each case.  We expect that the recovered cross section should be consistent with statistical noise once the limit is reached, as is clearly demonstrated. The inset in each panel of Fig.~\ref{fig:injsigmc} demonstrates how TS$_\text{max}$ depends on the injected cross section. 

While we show the representative cases for DM masses $m_\chi = 100$ and $1000$~GeV here, we find that signal injection tests on Monte Carlo are well-behaved for DM masses ranging from 10--1000~GeV.  The tests fail when the upper cutoff on the photon energy is $\gtrsim 100$~GeV most likely due to limited photon statistics.  For this reason, as well as the fact that the \texttt{p6v11} template should be used with caution at energies $\gtrsim 50$~GeV, we have restricted the photon energies to be below $\sim$50~GeV. \\

\begin{figure*}[t]
\centering
\includegraphics[width=0.48\textwidth]{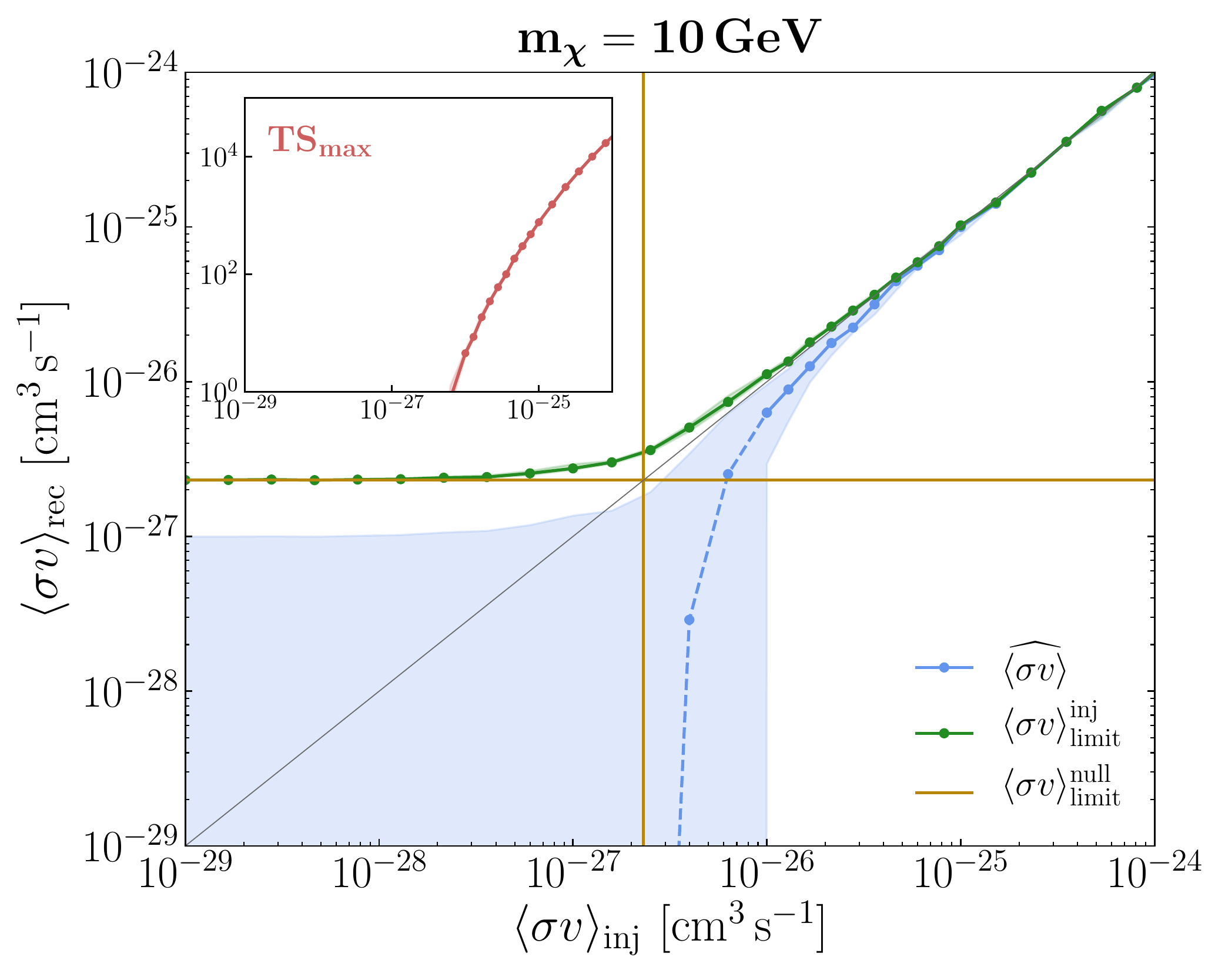} \hspace{4mm}
\includegraphics[width=0.48\textwidth]{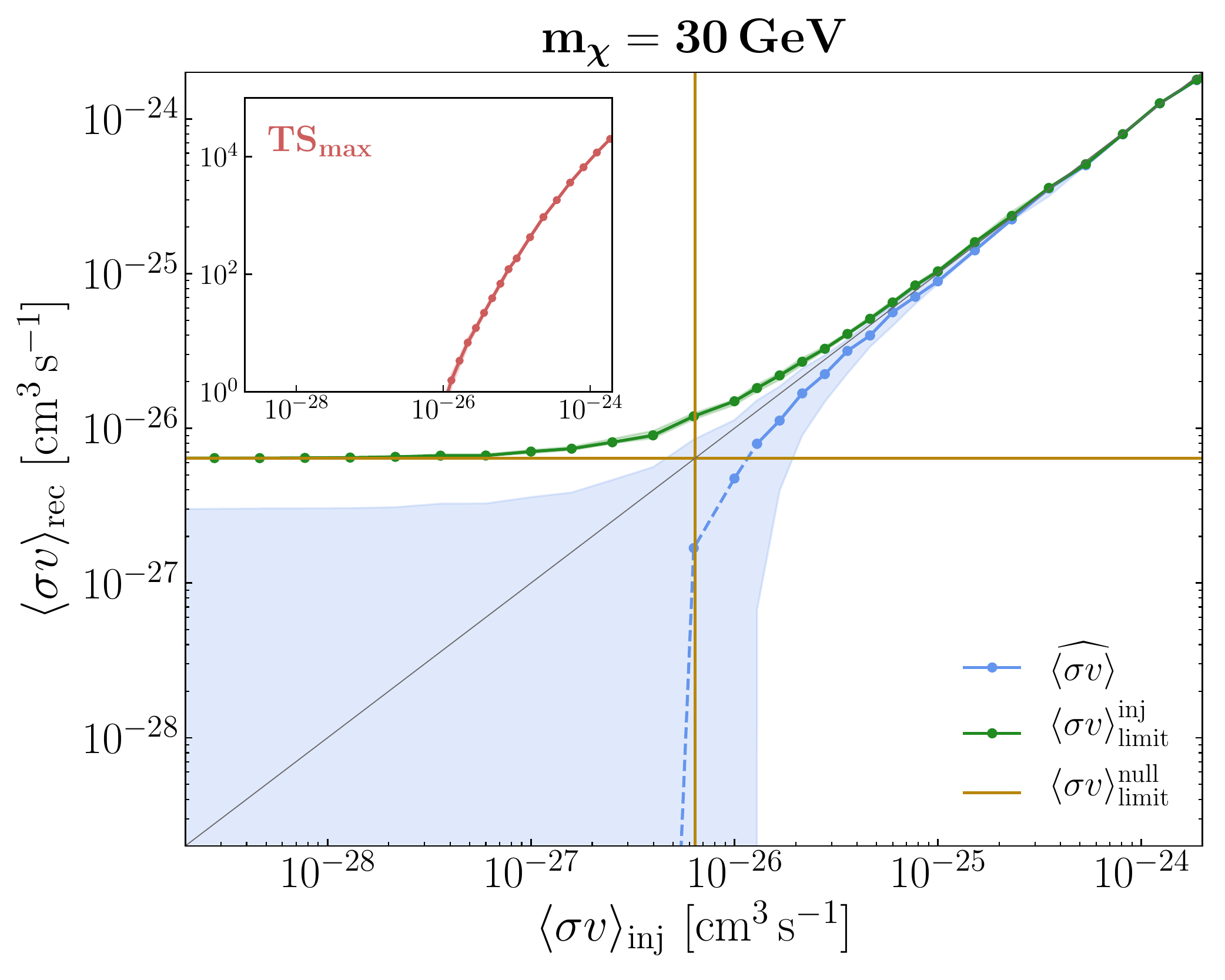} \hspace{4mm}
\caption{The same as Fig.~\ref{fig:injsigmc}, except for signal injected on data.  The left(right) panel corresponds to a 10(30)~GeV DM mass.  In this case, for each injected signal point, we create 10 realizations of simulated sky maps.}
\label{fig:injsigdata}
\end{figure*}

\noindent
\textbf{Signal injection on data.} We also perform a data-driven version of the signal injection tests, adding a Galactic DM signal for $b\bar b$ annihilation on top of the actual data and passing this through the analysis pipeline. We repeat this procedure for 10 sky map realizations.  This is a particularly important check at lower energies, where effects of point spread function (PSF) and foreground mis-modeling can lead to artificially strong limits for lower DM masses. Figure~\ref{fig:injsigdata} summarizes the results of the signal injection tests on data for DM masses of 10 and 30~GeV (left and right panel, respectively).  In each case, we see that the analysis would not exclude an injected DM signal. We restrict ourselves to energies $E_\gamma \gtrsim 0.8$~GeV to mitigate the effects of a significantly degraded PSF at even lower energies. We caution that while this procedure demonstrates that a signal would not be excluded under the null assumption on the data, it is still possible that mis-modeling effects can impact the final result for the lowest masses ($\sim$10 GeV).  This can be seen in the left panel of Fig.~\ref{fig:injsigdata} from the fact that the median recovered cross section associated with TS$_\text{max}$ (blue line) falls and becomes consistent with zero slightly above the null limit.  However, this small discrepancy occurs in the range where TS$_\text{max} \lesssim 1$. This is not an issue for higher masses; for example, for a DM mass of 30~GeV, the median recovered cross section associated with TS$_\text{max}$ is consistent with zero only for cross sections below the null limit, as shown in the right panel of Fig.~\ref{fig:injsigdata}. \\
\begin{figure*}[tb]
\centering
\includegraphics[width=0.48\textwidth]{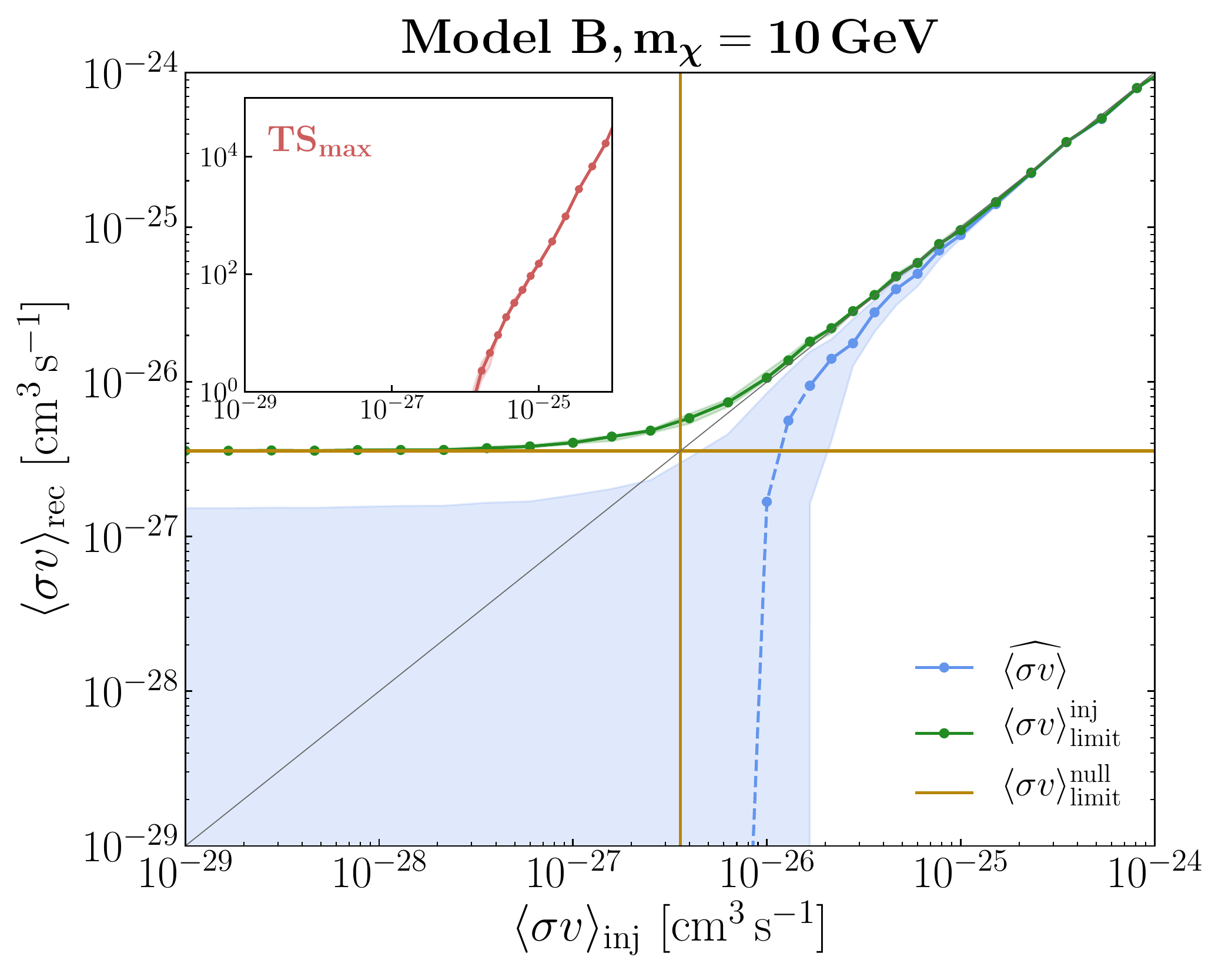} \hspace{4mm}
\includegraphics[width=0.48\textwidth]{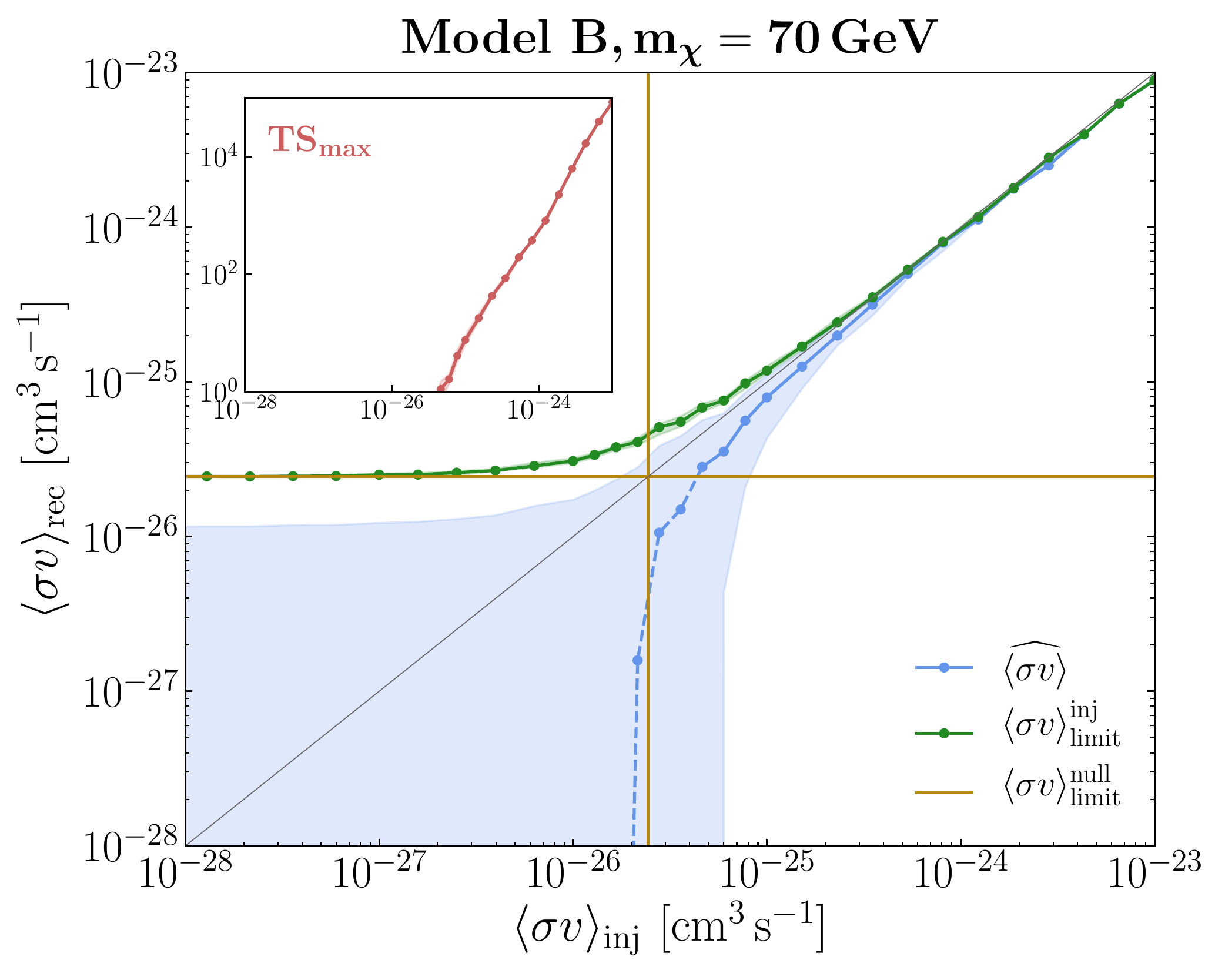} \hspace{4mm}
\caption{Signal injection test on data using the Model~B foreground template, assuming $m_\chi = 10$~GeV (left) and 70~GeV~(right).  Format as in Fig.~\ref{fig:injsigdata}. }
\label{fig:modelABCsiginj}
\end{figure*}

We have also performed signal injection tests using Model B to ensure the validity  of the recovered bounds.  The left panel of Fig.~\ref{fig:modelABCsiginj} shows the results of signal injection on data, as described in the previous section, for a DM mass of $m_\chi = 10$~GeV where foreground and PSF mis-modeling are likely to have the largest effect. We see that a putative DM signal would not be excluded by the analysis in this case. We also show results for an injected DM mass of $m_\chi = 70$~GeV in the right panel of Fig.~\ref{fig:modelABCsiginj}, corresponding to the value most consistent with the excess emission seen in Models~A and C. Again, we see that a potential DM signal would not be excluded in this case.

\section{Extended Results}
\label{sec:extended}

We consider several additional variations to the fiducial analysis, and summarize the results here:
\begin{itemize}

\item Although we presented results for DM annihilating into the $b\bar b$ and $\tau^+\tau^-$ final states in the paper, DM annihilation can proceed into a variety of Standard Model final states. In Fig.~\ref{fig:channels} (left), we reinterpret the main results of the fiducial study in terms of annihilation into additional final states. Broadly, the spectra of hadronic channels ($W^+W^-,~ZZ,~q\bar q,~c\bar c,~t\bar t$) are predominantly set by boosted $\pi^0$ decays, resulting in comparable final limits beyond the respective mass thresholds. Gamma-rays for the leptonic ($e^+e^-, \mu^+\mu^-$) channels predominantly arise from radiative decays and final-state radiation, resulting in somewhat weaker overall limits.  In each case, we assume 100\% branching fraction into the specified channel.  Note that we only model prompt gamma-ray emission and do not account for inverse-Compton or synchrotron radiation of the final state~\cite{Cirelli:2010xx}, which is relevant for the lighter leptonic channels. 

\item In addition to the $b\overline{b}$ and $\tau^+\tau^-$ cases considered in the paper, we summarize in Fig.~\ref{fig:channels} (right) constraints on other possible annihilation channels contributing to the GeV excess. We show our results for the $q\overline{q}, c\overline{c}, gg$ and $hh$ final states, spanning the range $\gamma=1.2$--1.3 for the inner slope of the NFW generalized profile (thick bands), along with the corresponding best-fit contours as found by~\cite{Calore:2014nla} assuming $\gamma=1.28$. We see that the $q\overline{q}$ and $hh$ explanations are robustly excluded by this analysis, while the $c\overline{c}$ and $gg$ explanations are put significantly under tension. We do not include annihilation channels that are already excluded at the 95\% confidence level by spectral fits to the \emph{Fermi} GeV excess emission~\cite{Calore:2014nla}.

\item Figure~\ref{fig:gce_bounds} of the paper demonstrates how the fiducial limit depends on the inner slope of the NFW profile.  We have additionally considered the Einasto and Burkert profiles, defined in Eq.~\ref{eq:Einasto} and \ref{eq:Burkert}.  The associated limits are shown in Fig.~\ref{fig:extensions}.  The Einasto limit (solid green) is a factor of $\lesssim1.6$ stronger than the fiducial case, while the Burkert limit is a factor of $\lesssim24(5)$ stronger(weaker) for $r_B = 0.5(10)$~kpc (dotted and dashed green, respectively).  	

\item We assumed a local DM density of $\rho(r_\odot) = 0.4$ GeV$\,$cm$^{-3}$ in the fiducial analysis, consistent with recent measurements~\cite{2015ApJ...814...13M, Sivertsson:2017rkp}. Other estimates in the literature, however, point to a value closer to $\rho(r_\odot) = 0.3$~GeV$\,$cm$^{-3}$ (see~\cite{Read:2014qva} and references therein). Repeating the analysis using this lower value, we find that the limit is $\lesssim1.8$ times weaker (solid blue line in Fig.~\ref{fig:extensions}).  We emphasize that the assumption made about the local DM density does not impact the conclusions drawn about the viability of the GeV excess, as the best-fit regions are similarly shifted to higher annihilation cross sections by roughly the same amount.

\begin{figure*}[t]
\centering
\includegraphics[width=0.47\textwidth]{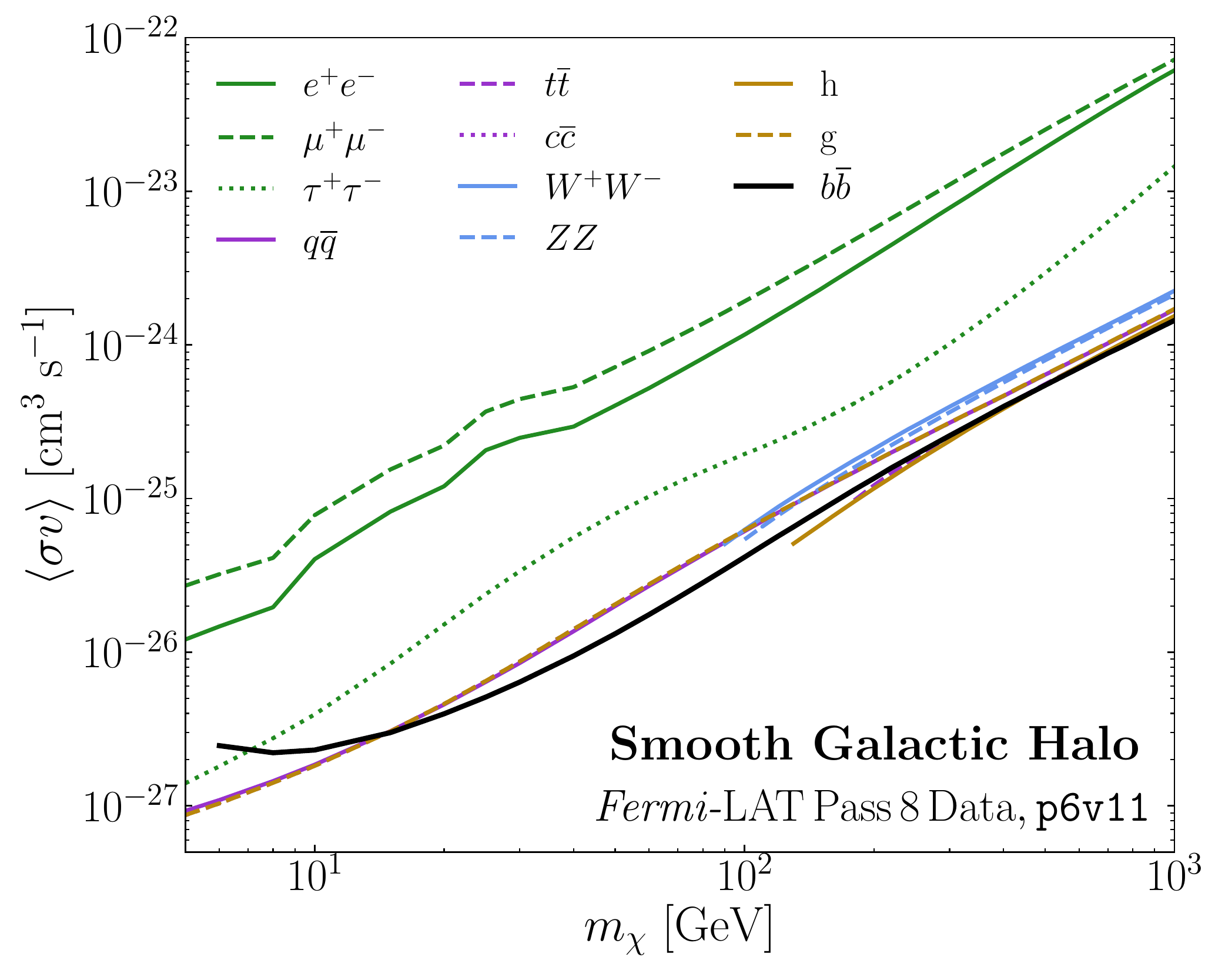} \hspace{4mm}
\includegraphics[width=0.47\textwidth]{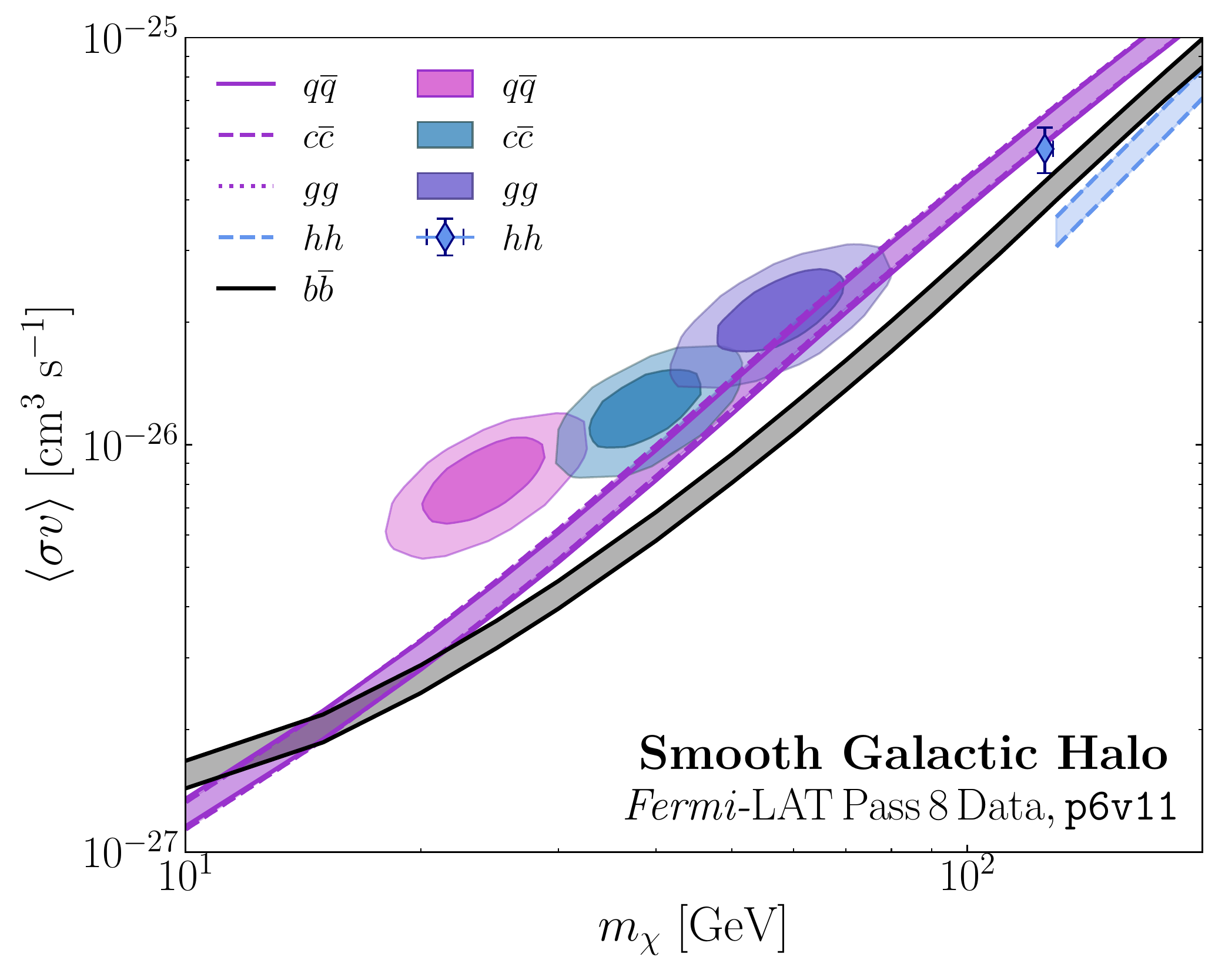} \hspace{4mm}

\caption{(Left) The 95\% confidence limit on dark matter of mass, $m_\chi$, annihilating with cross section, $\langle \sigma v \rangle$, in the smooth Galactic halo.  The limits are obtained following the fiducial analysis procedure described in the paper, but varying over the annihilation channel.    (Right) The 95\% confidence limits on dark matter annihilation into $b\bar b$ (fiducial), $q\bar q$, $c\bar c$, $gg$, and $hh$, varying over the inner slope, $\gamma$, of the generalized NFW density profile. The bands correspond to $\gamma$ values spanning $1.2$--$1.3$.  Note that the bands for $q\bar q$, $c\bar c$, and $gg$ fall essentially on top of each other.  The best-fit parameters for the $q\bar q$, $c\bar c$, and $gg$ channels, as obtained in~\cite{Calore:2014nla}, are indicated by the pink, teal, and purple 1$\sigma$/2$\sigma$ filled contours, respectively. The best-fit $hh$ value (and associated $1\sigma$ range) is indicated by the blue diamond~\cite{Calore:2014nla}.}
\label{fig:channels}
\end{figure*}

\item Our fiducial analysis does not account for potential emission from Loop~I in the Northern hemisphere.  As a proxy for this contribution, we include an additional isotropic template in the Northern hemisphere. Modeling this emission results in a slight improvement in the DM constraint by a factor of $\lesssim1.2$ (dashed purple line in Fig.~\ref{fig:extensions}), as expected because additional foreground components are accounted for.

\item In the fiducial study, the Northern and Southern lobes of the \emph{Fermi} bubbles are floated separately. We have verified that floating the Northern and Southern lobes together leave the limit unchanged. Figure~\ref{fig:extensions}~ shows what happens if the \emph{Fermi} bubbles are not included at all in the analysis. In this case, the limit worsens by a factor of $\lesssim6$ (solid gold line).  

\item In the fiducial study, all point sources were masked to 95\% containment in PSF, according to energy bin.  To estimate the effect of point-source mis-modeling, we increased the mask size to 99\% PSF containment; this results in a factor of $\lesssim1.5$ weakening of the fiducial limit (solid purple line in Fig.~\ref{fig:extensions}), likely due to the corresponding reduction in the effective size of the ROI.  

\begin{figure*}[t]
\centering
\includegraphics[width=0.47\textwidth]{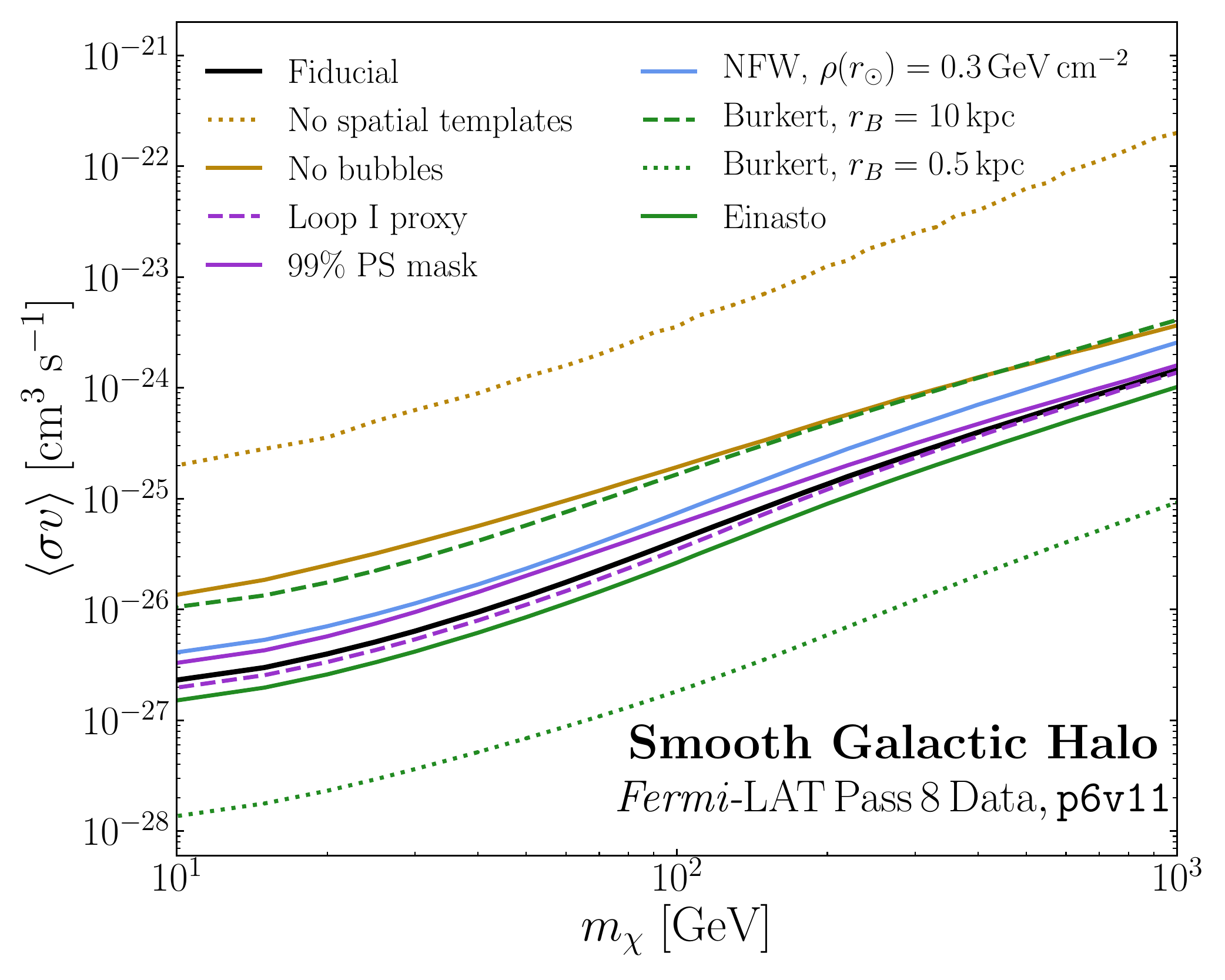} \hspace{4mm}

\caption{The 95\% confidence limits associated with variations to the fiducial analysis, as labeled in the legend and described in the text.  }
\label{fig:extensions}
\end{figure*}

\item The fiducial analysis takes full advantage of the spatial profiles of the expected DM emission and astrophysical components because we sum up the pixel-wise likelihoods. To quantify the gain from using spatial templates, we instead perform the fit using only the total expected number of counts from the DM signal and backgrounds within our ROI, and profile over the astrophysical nuisance parameters. The resulting limit (dotted gold line in Fig.~\ref{fig:extensions}) is several orders of magnitude weaker than the fiducial bound.

\item We show results obtained using the newer \texttt{p7v6} and \texttt{p8R2} diffuse models in Fig.~\ref{fig:p7_p8} (left and right panel, respectively).  As outlined in the paper, these models have large-scale residuals added back in to various extents, and as such are unsuitable for use in studying large-scale DM structures such as emission from the Galactic halo.  Indeed, in both cases, we observe significant over-subtraction for the fiducial ROI.  We emphasize that Fig.~\ref{fig:p7_p8} is included for illustration only and should be treated with caution.

\begin{figure*}[tb]
\centering
\includegraphics[width=0.47\textwidth]{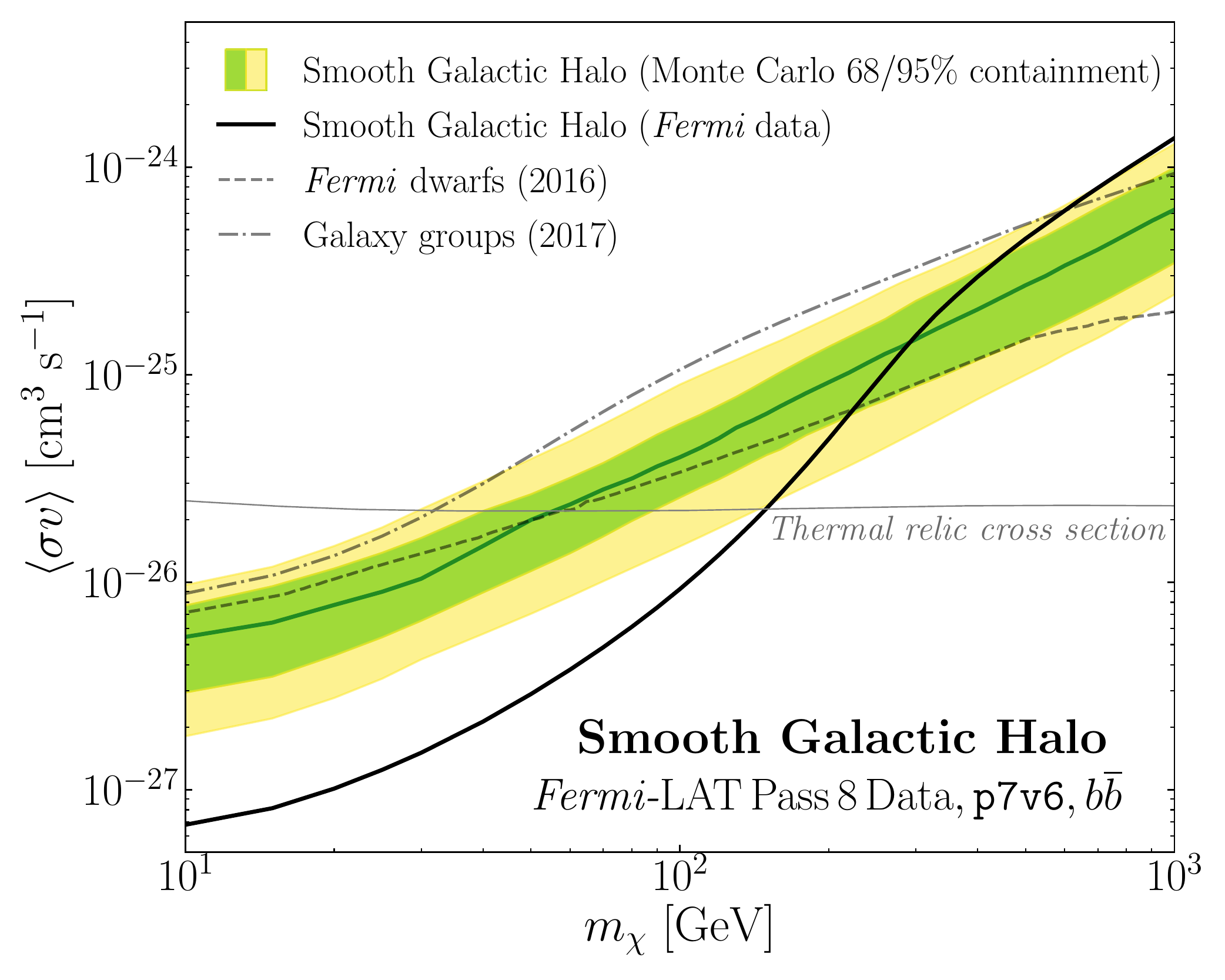} \hspace{4mm}
\includegraphics[width=0.47\textwidth]{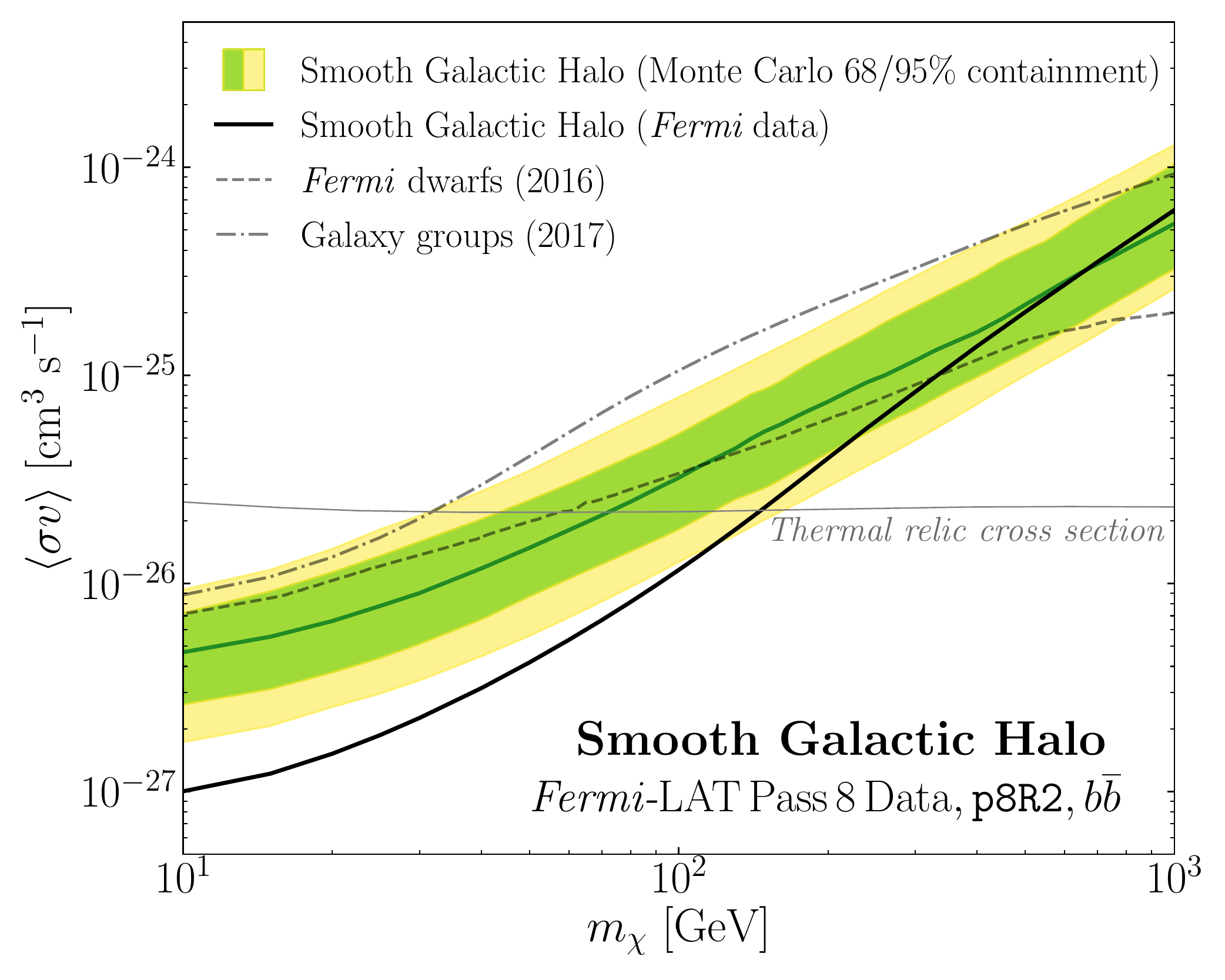} \hspace{4mm}

\caption{Similar to Fig.~\ref{fig:bounds} of the paper, except using the \texttt{p7v6} and \texttt{p8R2}  foreground models (left and right panel, respectively)~\cite{Acero:2016qlg}.  We only include these results for illustration as both of these foreground models are not appropriate for studies of diffuse DM signals, as discussed in the text.}
\label{fig:p7_p8}
\end{figure*}

\item Because the \emph{Fermi} bubbles are not accounted for when constructing the \texttt{p6v11} foreground model, one potential concern is the overestimation of the IC contribution. This could lead to inadequate modeling of the bubbles and potentially over-subtract a DM contribution, leading to an artificially strong limit. We show in Fig.~\ref{fig:bub_spectra} the energy spectra of the Northern (left) and Southern (right) lobes of the \emph{Fermi} bubbles as obtained from our analysis pipeline when using the various foreground models presented here. The spectra recovered when using \texttt{p6v11} are broadly similar to those obtained with Models A, B and C, underscoring the fact that the bubbles are adequately modeled in all four cases. We also show the bubbles spectra from~\cite{Fermi-LAT:2014sfa}, obtained for a slightly different ROI ($|b| > 10^\circ$ as opposed to $|b| > 20^\circ$), which are again similar to those derived in our analysis.

\begin{figure*}[htbp]
\centering
\includegraphics[width=0.91\textwidth]{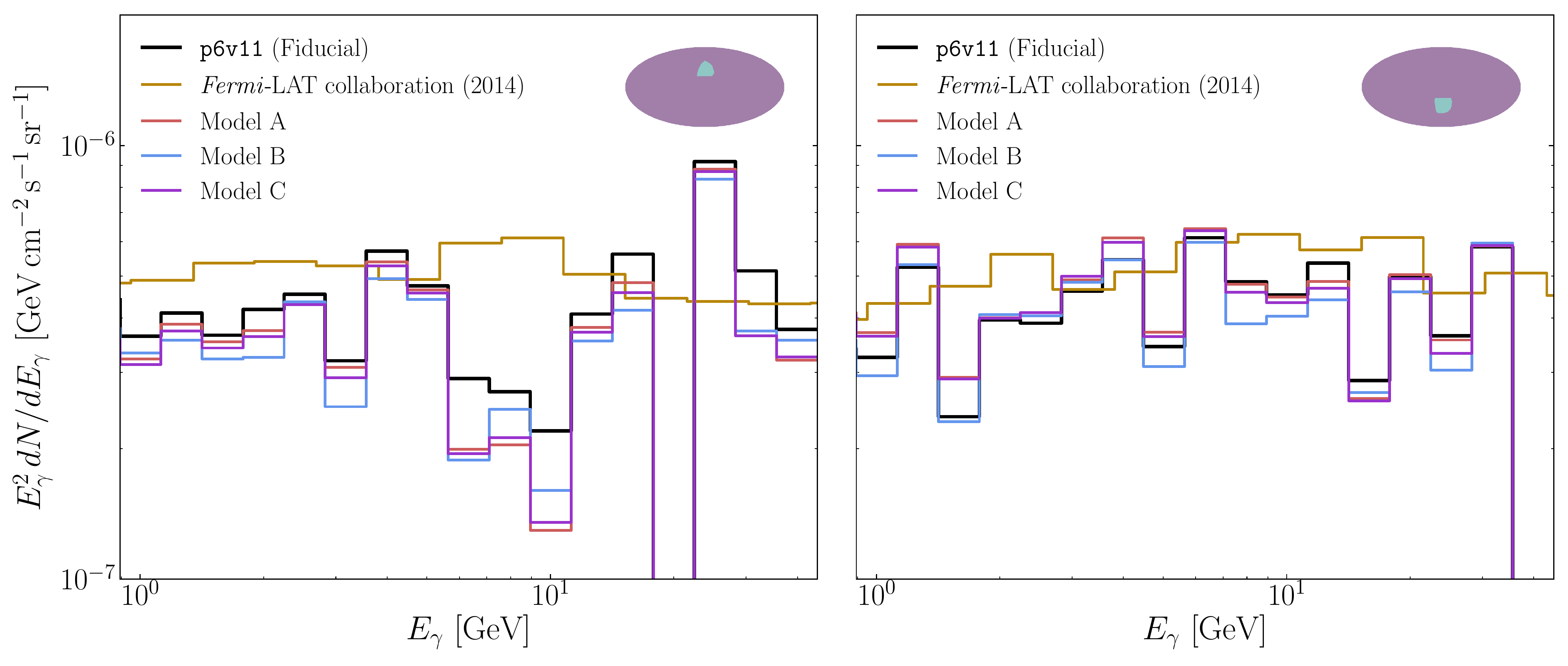} \hspace{4mm}
\caption{Recovered spectra, normalized to the corresponding bubbles region shown, for the Northern (left) and Southern (right) lobes of the \emph{Fermi} bubbles when analyzed with diffuse model \texttt{p6v11} as well as Models A, B and C. Our fiducial configuration was used to extract these spectra. The bubbles spectra obtained in~\cite{Fermi-LAT:2014sfa} are shown for comparison. Note that a slighty different ROI ($|b| > 10^\circ$ as opposed to $|b| > 20^\circ$) was used in that case. The energy $E_\gamma$ corresponds to the geometric mean of the energy bin edges.}
\label{fig:bub_spectra}
\end{figure*}

\item Given the importance of diffuse foreground modeling in the present study and the potential issues associated with a spectrally hard IC component in the \texttt{p6v11} model ~\cite{Calore:2014xka}, in Fig.~\ref{fig:difspec} we show the total energy spectra obtained for the \texttt{p6v11} model as well as those for Models A, B, and C in the eight radial slices considered in our study. We see that the spectra associated with \texttt{p6v11} (black line) are roughly consistent with the total spectra associated with Models A, B and C (red, blue and purple lines respectively).

\item Figure~\ref{fig:likelihoods} demonstrates the likelihood profiles for the fiducial analysis.  In general, there is very good agreement between the observed profile (black line) and the Monte Carlo expectation (blue band), in each energy bin.\\

\end{itemize}

\begin{figure*}[htbp]
\centering
\includegraphics[width=0.78\textwidth]{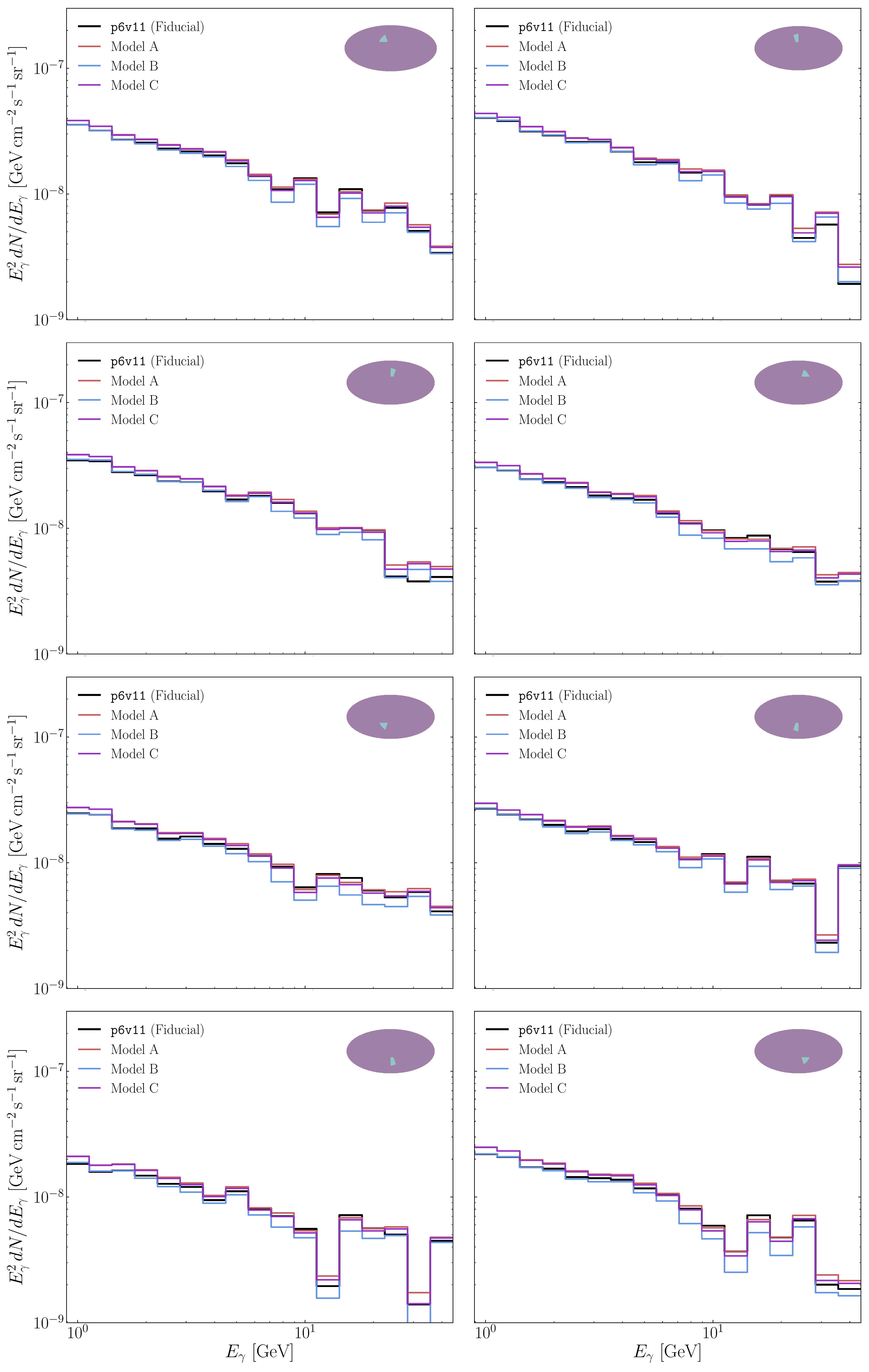} \hspace{4mm}
\caption{Energy spectra obtained for the \texttt{p6v11} model (black) as well as those for Models A, B, and C (red, blue and purple respectively) in the eight radial slices (shown as insets) considered in our study.}
\label{fig:difspec}
\end{figure*}

\begin{figure*}[htbp]
\centering
\includegraphics[width=\textwidth]{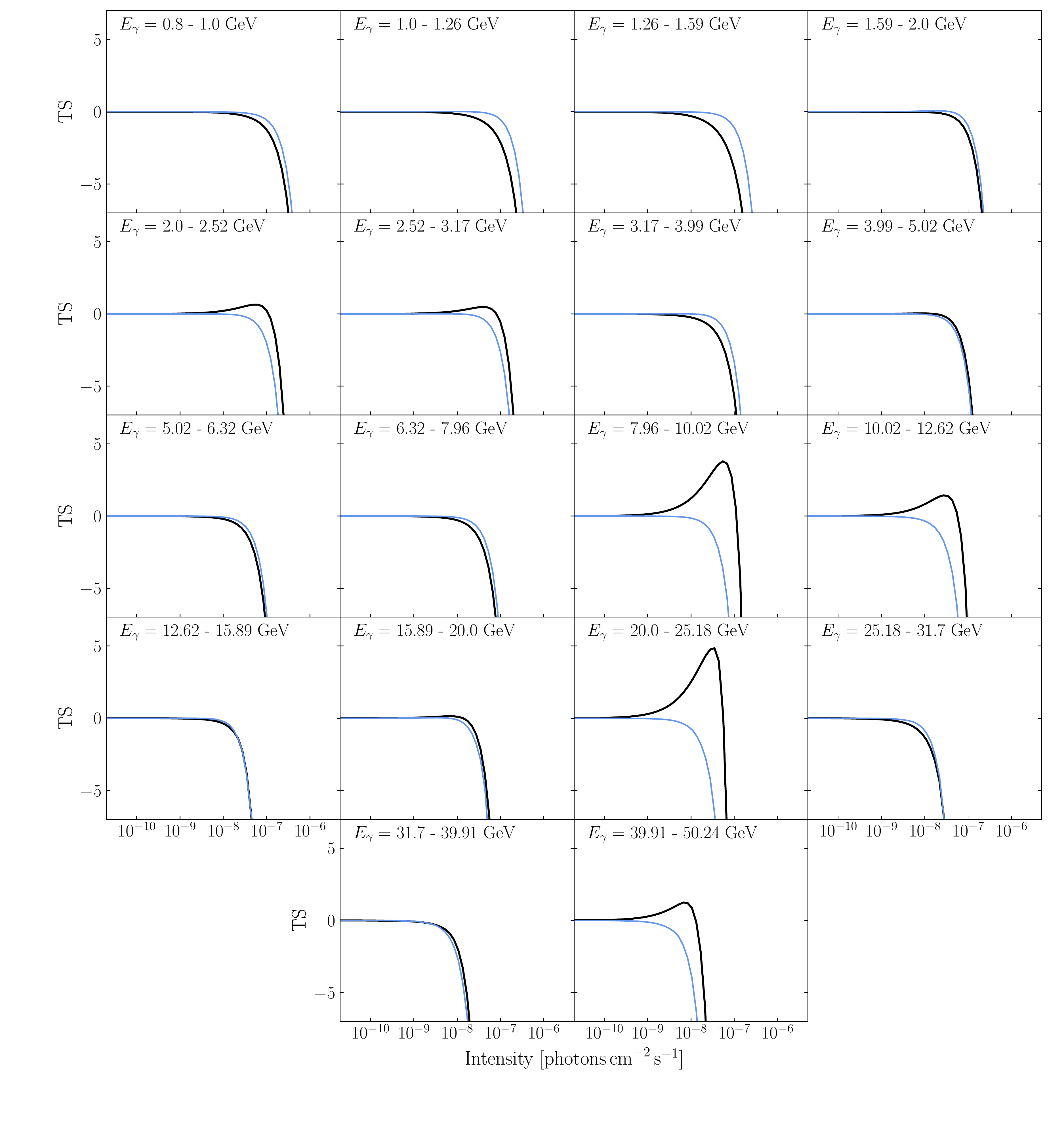} \hspace{4mm}
\caption{Likelihood profiles for the fiducial analysis (black lines), presented for each energy bin.  The darker(lighter) blue bands denote the 68(95)\% containment for the profiles, as determined from 100 Monte Carlo simulations.   }
\label{fig:likelihoods}
\end{figure*}

\end{document}